\documentclass{article}
\usepackage{graphicx}
\usepackage{epstopdf}
\usepackage{amsmath}
\usepackage{geometry}
\usepackage{amsfonts}
\usepackage{amssymb}
\usepackage{color}
\usepackage{lscape}
\usepackage[colorlinks=true,citecolor=blue]{hyperref}
\newcommand{\doublespacing}{\let\CS=\@currsize\renewcommand{\baselinesstrech}{2.0}\tiny\CS}
\begin{document}
	\newcommand{\bd}{\begin{document}}
		\newcommand{\ed}{\end{document}}
	\newcommand{\bc}{\begin{center}}
		\newcommand{\ec}{\end{center}}
	\newcommand{\bfr}{\begin{flushright}}
		\newcommand{\efr}{\end{flushright}}
	\newcommand{\lt}{\left}
	\newcommand{\rt}{\right}
	\newcommand{\vs}{\vspace}
	\newcommand{\hs}{\hspace}
	\newcommand{\beq}{\begin{equation}}
		\newcommand{\eeq}{\end{equation}}
	\newcommand{\lb}{\linebreak}
	\newcommand{\pb}{\pagebreak}
	\newcommand{\mb}{\makebox}
	\newcommand{\fb}{\framebox}
	\newcommand{\mc}{\multicolumn}
	\newcommand{\ben}{\begin{enumerate}}
		\newcommand{\een}{\end{enumerate}}
	\newcommand{\bit}{\begin{itemize}}
		\newcommand{\eit}{\end{itemize}}
	\newcommand{\oln}{\overline}
	\newcommand{\un}{\underline}
	\newcommand{\lefq}{\lefteqn}
	\newcommand{\ba}{\begin{array}}
		\newcommand{\ea}{\end{array}}
	\newcommand{\beqa}{\begin{eqnarray}}
		\newcommand{\eeqa}{\end{eqnarray}}
	\newcommand{\beqas}{\begin{eqnarray*}}
		\newcommand{\eeqas}{\end{eqnarray*}}
	\newcommand{\bfg}{\begin{figure}}
		\newcommand{\efg}{\end{figure}}
	\newcommand{\bds}{\begin{displaymath}}
		\newcommand{\eds}{\end{displaymath}}
	\newcommand{\btb}{\begin{tabbing}}
		\newcommand{\etb}{\end{tabbing}}
	\newcommand{\para}{\parallel}
	\newcommand{\pad}{\partial}
	\newcommand{\nn}{\nonumber}
	\newcommand{\la}{\leftarrow}
	\newcommand{\ra}{\rightarrow}
	\newcommand{\lgla}{\longleftarrow}
	\newcommand{\lgra}{\longrightarrow}
	\newcommand{\La}{\Leftarrow}\newcommand{\Ra}{\Rightarrow}
	\newcommand{\Lra}{\Leftrightarrow}
	\newcommand{\Lgla}{\Longleftarrow}
	\newcommand{\Lgra}{\Longrightarrow}
	\newcommand{\lan}{\langle}
	\newcommand{\ran}{\rangle}
	\renewcommand{\a}{\alpha}
	\renewcommand{\b}{\beta}
	\newcommand{\g}{\gamma}
	\newcommand{\G}{\Gamma}
	\renewcommand{\d}{\delta}
	\newcommand{\eps}{\epsilon}
	\newcommand{\Th}{\Theta}
	\newcommand{\s}{\sigma}
	\newcommand{\lam}{\lambda}
	\newcommand{\D}{\Delta}
	\newcommand{\ds}{\displaystyle}
	\newcommand{\vare}{E}
	\newcommand{\pr}{\prime}
	\newcommand{\ro}{\rho}
	\newcommand{\nab}{\nabla}
	\newcommand{\m}{\mu}
	\newcommand{\n}{\nu}
	\newcommand{\Sg}{\Sigma}
	\newcommand{\p}{\pi}
	\newcommand{\R}{I\!\!R}
	\newcommand{\om}{\omega}
	\newcommand{\Om}{\Omega}
	\newcommand{\ovra}{\overrightarrow}
	\newcommand{\ze}{\zeta}
	\newcommand{\vart}{\vartheta}
	\newcommand{\tri}{\triangle}
	\newcommand{\f}{\frac}
	\newcommand{\iny}{\infty}
	\newcommand{\pro}{\propto}
	\renewcommand{\arraystretch}{1.25}
	
	\bc {\huge Information theoretic measures of isotropic Dunkl oscillator in spherical coordinates} \ec
	\bc
	{\it
		Akash Halder \& Amlan K. Roy{\footnote{Email: akroy@iiserkol.ac.in, akroy6k@gmail.com}}\\Department of Chemical Sciences, Indian Institute of Science Education and Research (IISER) Kolkata, Mohanpur-741246, Nadia, WB, India.\\\&\\
		Debraj Nath{\footnote {Email: debrajn@gmail.com}} \\Department of Mathematics, Vivekananda College, 269 D.H. Road, Kolkata-700063, WB, India.
		
	} \ec
	
	\bc {\large {\un{\textbf{Abstract}}}} \ec
        An information theoretic analysis is done for the isotropic harmonic oscillator potential within the Dunkl-Schr\"odinger 
	framework in spherical coordinates. Starting from the \emph{exact} analytical eigensolution, various quantum information 
	measures such as Shannon entropy, R\'enyi information, Tsallis entropy are derived. Besides, their relative measures like 
	relative Shannon, relative R\'enyi, relative Tsallis as well as corresponding divergences (Jensen-Shannon, Jensen-R\'enyi, 
	Jensen-Tsallis) are also obtained. In order to get Shannon entropy, a novel factorization method is introduced.
	This is facilitated through the use of well-known weighted Lebesgue measure. The results from the Dunkl case agree exactly 
	with the non-Dunkl scenario, when Dunkl parameters vanish. The reflection operators and Dunkl parameters considerably 
	influence the above measures. These are portrayed in graphical forms.    \\	
	
	\date{\today}
	
	
	\section{Introduction}
	The study of deformation has some relevance in the realms of quantum algebra and groups in physics and mathematics. There is a particular interest in the Wigner-Dunkl formulation of quantum mechanics, originally pioneered in the 1950s \cite{wigner50}. There they investigated the generation of quantum mechanical observables from equations of motion. In accordance with the Wigner algebra (also called Wigner quantization method), a reflection operator $\widehat{R}$ (satisfying the relation $\widehat{R} f(x) = f(-x)$), multiplied by some constant, $\mu$ (termed as Wigner parameter) can be added in the regular position-momentum commutation relation giving: $[\widehat{x}, \widehat{p}]=i\hbar(1+ 2\mu \widehat{R})$. This novel algebra follows from a modification of the Boson algebra; it reduces to the latter when $\mu$ vanishes. Its connection to two-particle Calogero model has been discussed in the literature, where the Wigner parameter plays the role of Calogero coupling constant. In this scenario, relations like $\widehat{R} \widehat{x} = -\widehat{x}\widehat{R}, \ \widehat{R} \widehat{p} = -\widehat{p} \widehat{R}, \ \widehat{R} \widehat{R} =\textbf{I}$ hold, where \textbf{I} signifies the identity operator. The deformed algebra is not unique in position representation and a particular representation is through Dunkl operator \cite{dunkl2}: $\widehat{p}_x= \ds\frac{\hbar}{i} D_x= \ds\frac{\hbar}{i} \left[ \frac{d}{dx} + \frac{\mu}{x} (1-\widehat{R}) \right]$, where $\widehat{p}_x$ represents the ordinary momentum operator. It is expressed as a combination of differential and reflection operators. In one dimension, this can be written in standard form as $\ds D_x f(x) = \frac{d}{d x} f(x) + \frac{\mu}{x} (1-\widehat{R}) f(x)$.
	Due to the interplay between reflection operator and derivative, they play significant role in studying integrable systems and discrete symmetry quantum models. Furthermore it allows one to introduce both group symmetries and differentiation leading rise to newer algebraic structure, as well as spectral properties and creation of conserved quantities in a quantum system. Besides, the combination of $\mu$ parameter and $\widehat{R}$ operator ensures the wave function to be both continuous and discrete symmetries. A few variants of the operator exists, in the form of generalized Dunkl operator, Jacobi Dunkl operator etc. They arise in the context of reflection symmetries in multivariable polynomial, integral transform associated with root systems, complete set of quantum integrals, harmonic analysis, reflection symmetries in finite reflection group (also termed as finite Coexter group), quantum Calogero-Moser-Sutherland model and its generalization, anyons in (2+1) and (1+1) dimensions, para-fields and para-statistics, fractional statistics, conformal field theory, Dunkl-Laplacian and Dunkl transform, as well as various types of oscillators, Coulomb problem, relativistic problem etc. The result of Dunkl operator operating on a function depends on whether the function is even or odd.
	
	In the literature, a large number of works are reported for analytical solution of respective Dunkl-Schr\"odinger equation (DSE) in various model quantum systems, \emph{viz.}, in the particle in an infinite box in 1D box, wave functions are expressed in terms of Bessel functions and energies as zeros of Bessel function \cite{chung2019}. Eigenfunction and energies for DSE for harmonic oscillator (called Dunkl oscillator) in 1D are obtained in terms of $\nu$-deformed Hermite polynomial \cite{chung2019}. The ground and excited states of the oscillator were studied in momentum representation through Dunkl-Heisenberg relation \cite{chung2023}. For a generalized Dunkl oscillator in 1D, \emph{exact} eigenvalues, eigenfunctions as well as thermodynamic properties (Helmholtz free energy, mean energy, entropy) are reported in \cite{dong2021}. 
	The isotonic Mathews-Lakshmanan oscillator with Dunkl operator and position-dependent mass, does not admit bound state solution; however such states of even parity may occur if the isotonic term vanishes \cite{halberg2022}. 
	The Boltzmann factor and partition function of super statistics has been presented \cite{hassanabadi2021}. 
	
	The DSE has been discussed in 3D as well. The radial and angular equations of the spherical DSE are exactly solvable for free particle spherical wave, pseudo-harmonic oscillator and Mie-type potentials \cite{mota2022}. The analytical solutions \cite{genest2014a} of 3D isotropic harmonic oscillator are available in Cartesian (Hermite polynomial), cylindrical and spherical (radial, angular solutions in terms of Laguerre, Jacobi polynomials) coordinates. The system is maximally superintegrable, whose symmetries are obtained by Schwinger construction. For the \emph{generalized} DSE for harmonic oscillator is found by using generalized Dunkl derivative in Cartesian coordinates \cite{dong2023}, wherein the original symmetry of the Dunkl harmonic oscillator is \emph{broken} by generalized Dunkl derivative. The Dunkl-Klein-Gordon equation for the oscillator is separable in both Cartesian and spherical coordinates, where the eigenfunctions are written in terms of associated Laguerre and Jacobi polynomials. The general form of wave function in case of a DSE for the oscillator with time-dependent mass and frequency, in 1D and 3D has been derived by a Lewis-Riesenfeld method \cite{benchikha2024}. The problem has been solved analytically in higher dimensions in isotropic harmonic oscillator, Coulomb potential and pseudo-harmonic oscillator \cite{hamil2025}. The superintegrability and dynamical symmetry in Schwinger-Dunkl algebra has been analyzed as well \cite{genest2014a} in Dunkl oscillator and Dunkl-Coulomb problem. 
	
	

	The non-extensive one-parameter generalized entropy is called the Tsallis entropy $\mathcal{T}^{(\widetilde{\a})}$, which was introduced in 
	1988. It is used mainly in non-equilibrium statistical mechanics, as well as in ``long-range" interactions and complex systems. It has many applications in physics astrophysics. In the limiting case $\widetilde{\a}\rightarrow1$, $\mathcal{T}^{(\widetilde{\a})}$ gives $\mathcal{S}$. In 1948 \cite{shannon}, it was defined for the measure of uncertainty or randomness in a quantum state. It is well-known that, R\'enyi entropy $\mathcal{R}^{(\widetilde{\a})}$ introduces a weight parameter $\widetilde{\a}$ of probability distribution of a state \cite{renyi}. In the limiting case of $\widetilde{\a}\rightarrow1$ it gives $\mathcal{S}$, when $\widetilde{\a}\rightarrow0$ it gives Hartly entropy and for $\widetilde{\a}\rightarrow\infty$ it gives min-entropy. In density functional theory, the Thomas-Fermi kinetic energy functional can be derived from the maximum $\mathcal{R}$ principle for $\alpha=\frac{5}{3}$, whereas Dirac exchange energy is obtained from $\mathcal{R}$ of order $\widetilde{\a}=\frac{4}{3}$. The order-2 $\mathcal{R}^{(\widetilde{\a})}$ and the disequilibrium $\mathcal{D}$ or information energy are connected by the relation $\mathcal{R}^{(2)}=-\ln[\mathcal{D}]$. In quantum chemistry, $\mathcal{D}$ is identified as self similarity. Moreover, the structural entropy is defined by $\mathcal{R}$ as $S_{str}=\mathcal{R}^{(1)}-\mathcal{R}^{(2)}=\mathcal{S}-\mathcal{R}^{(2)}$. The linear entropy $1-\mathcal{D}$ is defined as the measure of impurity of a quantum system.  
	
	In this communication, we are interested in the information theoretical analysis of 3D Dunkl oscillator, for which no results are available, to the best of our knowledge. Recently a detailed work in respective 1D oscillator has been undertaken \cite{nath2024}. At first, in Sec.~\ref{sec2.information}, we derive the spectrum of eigenfunctions and eigenvalues of 3D Dunkl oscillator, which are expressed in terms of Jacobi polynomial. Then we obtain expectation values of $\langle r^2 \rangle$, $\langle r^{-2} \rangle$, uncertainties $\Delta r$, $\Delta p_r$ as well as information measures, such as R\'enyi entropy ($\mathcal{R}$), Tsallis entropy ($\mathcal{T}$) and Shannon entropy ($\mathcal{S}$), in Sec.~\ref{sec3.infor}. Next Sec.~\ref{sec4.relative} provides a bunch of relative information measures of two density functions, such as relative Shannon entropy and generalized Jensen-Shannon divergence (GJSD), relative R\'enyi entropy and generalized Jensen-R\'enyi divergence (GJRD), relative Tsallis entropy and Jensen-Tsallis divergence (JTD). The effect of Dunkl parameter and reflection operators on information theoretic measures are investigated. The analytic expressions are offered here for the first time, with accompanying figures in representative cases. A few conclusions are drawn in Sec.~\ref{sec5.con}.   
	\section{Information theoretical measures of the Dunkl oscillator}\label{sec2.information}
	\subsection{Introduction and background, Preamble}
	The time-independent DSE in 3D spherical polar coordinates is defined as,  
	$\left[-\frac{\hbar^2}{2\mu}\nabla_D^2+V(r)\right]\psi(r)=E\psi(r).$
	If $V(r)$ is a central potential, the Dunkl-Laplacian is defined by $\nabla_D^2=M_r-\frac{L_D^2}{\hbar^2r^2}$, where $L_D^2=-\hbar^2\left(N_{\theta}+\frac{1}{\sin^2\theta}B_{\phi}\right)$ is 
	the angular Dunkl momentum operator and the operators $M_r$, $N_{\theta}$, $B_{\phi}$ are defined in \cite{genest2014a,arxiv.dn} as:
	\beq
	\ba{ll}
	M_r&=\ds\frac{1}{r^{2a}}\frac{\partial}{\partial r}\left(r^{2a}\frac{\partial}{\partial r}\right),~a=1+\mu_x+\mu_y+\mu_z,\\
	N_{\theta}&=\ds\frac{1}{\sin\theta}\frac{\partial}{\partial\theta}\left(\sin\theta\frac{\partial}{\partial\theta}\right)+2\left[(\mu_x+\mu_y)\cot\theta-\mu_z\tan\theta\right]\frac{\partial}{\partial\theta}-\frac{\mu_z}{\cos^2\theta}(1-\widehat{R}_z),\\
	B_{\phi}&=\ds\frac{\partial^2}{\partial\phi^2}-2\left(\mu_x\tan\phi-\mu_y\cot\phi\right)\frac{\partial}{\partial\phi}-\frac{\mu_x}{\cos^2\phi}(1-\widehat{R}_x)-\frac{\mu_y}{\sin^2\phi}(1-\widehat{R}_y).
	\ea 
	\eeq 
	The reflection operators act in the spherical coordinates as: $\widehat{R}_x\psi(r,\theta,\phi)=\psi(r,\theta,\pi-\phi)$, $\widehat{R}_y\psi(r,\theta,\phi)=\psi(r,\theta,-\phi)$, $\widehat{R}_z\psi(r,\theta,\phi)=\psi(r,\pi-\theta,\phi)$, whereas it acts in Cartesian coordinate as follows: $\widehat{R}_xf(x,y,z)=f(-x,y,z)$, $\widehat{R}_yf(x,y,z)=f(x,-y,z)$, $\widehat{R}_zf(x,y,z)=f(x,y,-z)$. Here $\mu_x,\mu_y,\mu_z$ are called Dunkl parameters. The $\phi$-angular solution  $G (\phi)$ is found to be explicitly dependent on parity eigenvalues, $s_1$, $s_2$ of $\widehat{R}_x, \widehat{R}_y$. Likewise, $\theta$ angular wave function $H(\theta)$ depends explicitly on $\widehat{R}_z$ operator eigenvalue, $s_3$, while implicitly depending on $s_1,s_2$. The radial wave function $R_{n,\ell,m}$ implicitly depends on $s_1, s_2, s_3$.
	In this work, we have considered the isotropic harmonic oscillator potential: 
		$V(r)=\frac{1}{2} \mu \om^2r^2$, 
	where $\omega$ is the frequency of oscillator. 
	Then the solution of DSE can be expressed as: $\psi(r,\theta,\phi)=\frac{R(r)}{r^a}H(\theta)G(\phi)$, where 
	\beq
	\ba{ll}
	G_m^{(s_1,s_2)}(\phi)=N_{m}^{(\phi)}(1-p)^{\nu}(1+p)^{\xi}\,P_{m}^{(\a,\b)}(p),~p=\cos2\phi ,\\
	H_{\ell,m}^{(s_3)}(\theta)=N_{\ell,m}^{(\theta)}(1-q)^{\sigma}(1+q)^{\varsigma}\,P_{\ell}^{(\g,\d)}(q),~q=\cos2\theta, \\
	R_{n,\ell,m}(r)=N_{n,\ell,m}^{(r)}\,r^{L+1}e^{-\frac{\lam}{2}r^2}\,{}_1F_1\left(-n,L+\frac{3}{2},\lam r^2\right),
	\ea
	\eeq
	with 
	\beq
	\ba{llll}\label{sigma.varsigma}
	\a=2\nu+\mu_y-\frac{1}{2},~\b=2\xi+\mu_x-\frac{1}{2},&	\nu=\frac{1-s_2}{4},~ \xi=\frac{1-s_1}{4},\\
	\g=2\sigma+\mu_x+\mu_y,~\d=2\varsigma+\mu_z-\frac{1}{2},&	\sigma=\frac{1}{2}\left[-\mu_x-\mu_y+\sqrt{(\mu_x+\mu_y)^2-\eps^{(1)}_m}\right],~\varsigma=\frac{1-s_3}{4} \\
    \lam=\frac{\mu \omega}{\hbar},~ a=\mu_x+\mu_y+\mu_z+1&L=-\frac{1}{2} + \sqrt{\left(a-\frac{1}{2}\right)^2-\eps^{(2)}_{\ell,m}}.
	\ea
	\eeq 
	The separation constants are defined by:
	\beq
	\ba{ll}
	\eps^{(1)}_{m}=-4(\nu+\xi)(\nu+\xi+\mu_x+\mu_y)-4m(m+\a+\b+1), \\
	\eps^{(2)}_{\ell,m}=-4\ell(\ell+\g+\d+1)-4(\sigma+\varsigma)\left(\sigma+\varsigma+\mu_x+\mu_y+\mu_z+\frac{1}{2}\right).
	\ea
	\eeq
	In the above, $P_{m}^{(\a,\b)}(x)$ and ${}_1F_1\left(-n,L+\frac{3}{2},\lam r^2\right)$ denote Jacobi polynomial and hyper-geometric function respectively. 
	The normalization constants, 
	$N_m^{(\phi)}=\left[\frac{m!(\a+\b+2m+1) \G[\a+\b+m+1]}{2^{2\nu+2\xi+1} \G[\a+m+1] \G[\b+m+1]}\right]^{\frac{1}{2}}$,
	$N_{\ell,m}^{(\theta)}=\left[\frac{\ell!(\g+\d+2\ell+1) \G[\g+\d+\ell+1]}{2^{2\sigma+2\varsigma} \G[\g+\ell+1] \G[\d+\ell+1]}\right]^{\frac{1}{2}}$, and $N_{n,\ell,m}^{(r)}= \left[ \frac{2 \lambda^{L+\frac{3}{2}} ~ \G\left(n+L + \frac{3}{2} \right)}{n!~ \Gamma\left(L + \frac{3}{2}\right)^2} \right]^{\frac{1}{2}}$
	are obtained from orthogonality conditions: 
	$\int_0^{2\pi}G_{m}^{(s_1,s_2)}(\phi)G_{m'}^{(s'_1,s'_2)}(\phi)d\chi_\phi=\delta_{m,m'}\delta_{s_1,s_1'}\delta_{s_2,s_2'}$, $\int_0^{\pi}H_{\ell,m}^{(s_3)}(\theta)H_{\ell',m'}^{(s_3')}(\theta)d\chi_\theta=\delta_{\ell,\ell'}\delta_{s_3,s_3'}$,  
	$\int_0^{\infty}\frac{R_{n,\ell,m}(r)R_{n',\ell,m}(r)}{r^{2a}}d\chi_3=\delta_{n,n'}$
	where the measures are given by: $d\chi_\phi=|\cos\phi|^{2\mu_x}|\sin\phi|^{2\mu_y}d\phi$, $d\chi_\theta=|\sin\theta|^{2\mu_x+2\mu_y}|\cos\theta|^{2\mu_z}\sin\theta\,d\theta$ and $d\chi_3=r^{2a}dr$.
	
	The solution $G(\phi)$ can be written in an alternative form as:
	$G_m^{(s_1,s_2)}(\phi)=N_{m}^{(\phi)}(1-p)^{\nu}(1+p)^{\xi}\,P_{m-\nu-\xi}^{(\a,\b)}(p)$, with the corresponding separation constant depending on $m$ as: $\eps^{(1)}_{m}=-4m(m+\mu_x+\mu_y)$,  where $m=0$, only if $s_1=s_2=1$; $m\in\mathbb{N}$ if $s_1s_2=1$, and $m=\frac{1}{2}+k$, for some $k\in\mathbb{N}\cup\{0\}$ for every pairs  of $s_1,s_2$ such that $s_1s_2=-1$.
	Similarly, an alternate expression of $H (\theta)$ is given by: $H_{\ell,m}^{(s_3)}(\theta)=N_{\ell,m}^{(\theta)}(1-q)^{\sigma}(1+q)^{\varsigma}\,P_{\ell-\varsigma}^{(\g,\d)}(q)$, with $\eps^{(2)}_{\ell,m}=-4(\ell+m)\left(\ell+m+\mu_x+\mu_y+\mu_z+\frac{1}{2}\right) $, where $\ell\in\mathbb{N}\cup\{0\}$, if $s_3=1$, and $\ell=\frac{1}{2}+k$, for some $k\in\mathbb{N}\cup\{0\}$, if $s_3=-1$. Therefore, one sees that, $G(\phi)$ function is directly affected by $\widehat{R}_x, \widehat{R}_y$, while 
	$H(\theta)$ function is directly impacted by $\widehat{R}_z$ and indirectly by $\widehat{R}_x, \widehat{R}_y$. Finally, the eigenvalue of angular momentum 
	operator $L_D^2$ is $-\hbar^2\eps_{\ell,m}^{(2)}$. 


	
	In the usual Schr\"odinger system (ordinary derivative (OD)), the vibrational energy is obtained as: $E_{n,\ell,m}^{(OD)}=\hbar \omega(2n+\ell+\frac{3}{2})$, whereas in the Dunkl scenario (with Dunkl derivative (DD)), it is found as: $E_{n,\ell,m}^{(DD)}=\hbar \omega(2n+L+\frac{3}{2})=E_{n,\ell,m}^{(OD)}=\hbar \omega(2n+L+\frac{3}{2})$ if $\ell=-\frac{1}{2} + \sqrt{\left(a-\frac{1}{2}\right)^2-\eps^{(2)}_{\ell,m} }$. Thus, oscillator energy depends on $\mu_x, \mu_y, \mu_z$. Since, the energy $E_{n,\ell,m}^{(OD)}$ is degenerate with respect to $m$, one can find degenerate energy $E_{n,\ell,m}^{(DD)}$ in $m$ as well as in $\ell$. Moreover, one can find a set of quantum numbers $n,\ell,m$ for which $E_{n,\ell,m}^{(OD)}=E_{n,\ell,m}^{(DD)}$.
	In the proceeding sections, some information measures, such as linear information, relative entropy and divergence in DS system are considered. 

	\subsection{Linear information theoretic measures}\label{sec3.infor}
	The joint density function $\rho_{n,\ell,m}^{(s_1,s_2,s_3)}(\mathbf{r})$ can be expressed as $\rho_{n,\ell,m}^{(s_1,s_2,s_3)}(\mathbf{r})=\rho_{n,\ell,m}^{(r)}\rho_{\ell,m}^{(\theta,s_3)}\rho_{m}^{(\phi,s_1,s_2)}$, where $\rho_{n,\ell,m}^{(r)}=\frac{R_{n,\ell,m}^2(r)}{r^{2a}}$, $\rho_{\ell,m}^{(\theta,s_3)}=\left[H_{\ell,m}^{(s_3)}(\theta)\right]^{2}$, $\rho_{m}^{(\phi,s_1,s_2)}=\left[G_{m}^{(s_1,s_2)}(\phi)\right]^{2}$ are radial and angular ($\theta$ and $\phi$) density functions defined over the domains $r\in[0,\infty)=\mathbb{R}^*$, $\theta\in[0,\pi]$ and $\phi\in[0,2\pi]$. The density functions are orthogonal with respect to weighted Lebesgue measures, $d\chi_r$, $d\chi_{\theta}$, $d\chi_{\phi}$ respectively and form three weighted $\xi\left(\mathbb{R}^*,d\chi_r\right)$, $\xi\left([0,\pi],d\chi_{\theta}\right)$, $\xi\left([0,2\pi],d\chi_{\phi}\right)$ spaces \cite{vinet2013}. The composite (total) density functions are orthogonal with respect to the weighted Lebesgue measure $d\chi=d\chi_1d\chi_2d\chi_3$ and the corresponding weighted space is $\xi\left(\mathbb{R}^*\times[0,\pi]\times[0,2\pi],\,d\chi\right)$. Note that the Dunkl region is defined by $\{(\mu_x,\mu_y,\mu_z): \mu_x, \mu_y, \mu_z>-\frac{1}{2}\}$, while the Dunkl-plane segment by: $\mu_x+\mu_y+\mu_z=0$.
	
	The expectation $\left\langle r^j\right\rangle_{n,\ell,m}$ can be written as,  
	\beq \label{exprj}
	\ba{ll}
	\left\langle r^j\right\rangle_{n,\ell,m}&=\ds\int_0^{\infty} r^jR^2_{n,\ell,m}(r)dr=\ds\frac{ [N_{n,\ell,m}^{(r)}]^2}{2\lam^{ L + \frac{j+3}{2}}}\mathcal{A}_n^{(L + \frac{3}{2})}\left(1,L + \frac{j+3}{2},1\right),~j>-3
	\ea
\eeq
where $\mathcal{A}_{n}^{(\a)}(a,b,c)=\int_0^{\infty} e^{-as} s^{b-1}\left[{}_1F_1\left(-n,\a,s\right)\right]^{2c}ds$, which is defined analytically for $c\in\mathbb{N}$ as: 
\beq 
\mathcal{A}_{n}^{(\a)}(a,b,c)=\ds \sum_{i=0}^{2nc} \frac{(2c)! \widetilde{B}_{i+2c,2c} \left( c^{(n,\a)}_0,\cdots,(i+1)! c^{(n,\a)}_{i}\right)\Gamma\left(b + i\right)}{(i + 2c)!a^{b+i}},
\eeq 
with $c_i^{(n,\a)}=\frac{(-n)_i}{(\a)_i\,i!}$, for $i\le n$ and $c_i^{(n,\a)}=0$, for $i>n$. Here $\widetilde{B}_{m,\ell}(x_1,x_2,\dots,x_{m-l+1})$ denotes the Bell polynomial \cite{riordan1980}. When $j\le3$, one can find the numerical value of $\left\langle r^j\right\rangle_{n,\ell,m}$.

	Then the standard deviation $\left(\Delta r\right)_{n,\ell,m}=\sqrt{\left\langle r^2\right\rangle_{n,\ell,m}-\left\langle r\right\rangle_{n,\ell,m}^2}$ of radial operator can be expressed as, 
	\beq
	\left(\Delta r\right)_{n,\ell,m}=\ds \left[\frac{ [N_{n,\ell,m}^{(r)}]^2}{2\lam^{ L + \frac{5}{2}}}\mathcal{A}_n^{(L + \frac{3}{2})}\left(1,L + \frac{5}{2},1\right) -\frac{ [N_{n,\ell,m}^{(r)}]^4}{4\lam^{ 2L + 4}}\left\{\mathcal{A}_n^{(L + \frac{3}{2})}\left(1,L + 2,1\right) \right\}^2\right]^{1/2}.
	\eeq
	The radial momentum operator, $p_r$, in DSE system is defined by: $p_r=\ds-i\hbar\left(\frac{\partial}{\partial r}+\frac{a}{r}\right)$, satisfying the commutation relation $[r,p_r]=i\hbar$ and $p_r^2=-\hbar^2\left(M_r+\frac{a(a-1)}{r^2}\right)$. The expectations of $p_r$ and $p_r^2$ can be expressed as,
	\beq
	\ba{ll}
	\left\langle p_r\right\rangle_{n,\ell,m} &=0,\\
	\left\langle p_r^2\right\rangle_{n,\ell,m}&
	=2\mu E_{n,\ell,m}+\hbar^2(a(a-1)-\eps^{(2)}_{\ell,m})\left\langle\frac{1}{r^2 }\right\rangle_{n,\ell,m}-\mu^2\om^2\left\langle r^2\right\rangle_{n,\ell,m},
	\ea
	\eeq
	where $\left\langle r^2 \right\rangle$ and $\left\langle r^{-2} \right\rangle$ are obtained from Eq.~(\ref{exprj}).
	Then the RMS of radial momentum operator $p_r$ can be expressed as, 
	$\left(\Delta p_r\right)_{n,\ell,m}=\sqrt{\left\langle p_r^2\right\rangle_{n,\ell,m}-\left\langle p_r\right\rangle_{n,\ell,m}^2}= \sqrt{\left\langle p_r^2\right\rangle_{n,\ell,m}}$.
	Thus, the product of RMS of $r$ and $p$ is found to satisfy the following Heisenberg uncertainty relation 
	$(\Delta r)_{n,\ell,m}(\Delta p_r)_{n,\ell,m}\ge \frac{\hbar}{2}$.
	
	\subsubsection{Entropic moment, R\'enyi and Tsallis entropies} 
	The entropic moment of the density function $\rho_{n,\ell,m}$ is given by:
	\beq
	\mathcal{I}_{n,\ell,m}^{(s_1,s_2,s_3,\widetilde{\a})}=\ds\int[\rho_{n,\ell,m}^{(s_1,s_2,s_3)}(\mathbf{r})]^{\widetilde{\a}}d\chi=\mathcal{I}_{n,\ell,m}^{(r,\widetilde{\a})}\mathcal{I}_{\ell,m}^{(\theta,\widetilde{\a})}\mathcal{I}_m^{(\phi,\widetilde{\a})},
	\eeq 
	where $\mathcal{I}_{n,\ell,m}^{(r,\widetilde{\a})}$, $\mathcal{I}_{\ell,m}^{(\theta,\widetilde{\a})}$, $\mathcal{I}_{m}^{(\phi,\widetilde{\a})}$ are individual entropic moments of marginal density functions, given by: 
	$\mathcal{I}_{n,\ell,m}^{(r,\widetilde{\a})}=\int_0^{\infty}\left[\rho_{n,\ell,m}^{(r)}\right]^{\widetilde{\a}}d\chi_r$, $\mathcal{I}_{\ell,m}^{(\theta,s_3,\widetilde{\a})}=\int_0^{\pi}\left[\rho_{\ell,m}^{(\theta,s_3)}\right]^{\widetilde{\a}}d\chi_{\theta}$ and $\mathcal{I}_m^{(\phi,s_1,s_2,\widetilde{\a})}=\int_0^{\pi}\left[\rho_{m}^{(\phi,s_1,s_2)}\right]^{\widetilde{\a}}d\chi_{\phi}$ for $\widetilde{\a}>0,\ne 1$.
	Therefore, $\left\langle A\right\rangle_{n,\ell,m}^{(s_1,s_2,s_3,1)}=A$, if $A\in\mathbb{C}$, a constant. The entropic moments of positive integral order can be expressed as, 
	\beq
	\left.\ba{ll}
	\mathcal{I}_{n,\ell,m}^{(r,\widetilde{\a})}
	&=\ds\frac{[N_{n,\ell,m}^{(r)}]^{2\widetilde{\a}}}{2 \lam^{\widetilde{\a}(L+1) +\frac{1}{2}}}\mathcal{A}_{n}^{(L+\frac{3}{2})}\left(\widetilde{\a},\widetilde{\a}(L+1) -a(\widetilde{\a}-1)+ \frac{1}{2},\widetilde{\a}\right) ,\\
	\mathcal{I}_{\ell,m}^{(\theta,s_3,\widetilde{\a})}
	&=\ds\frac{[N_{\ell,m}^{(\theta)}]^{2\widetilde{\a}}}{2^{\mu_x+\mu_y+\mu_z+\frac{1}{2}}}\mathcal{B}_{\ell}^{(\g,\d)}\left(\mu_x+\mu_y+2\widetilde{\a} \sigma+1,\mu_z+2\widetilde{\a} \varsigma+\frac{1}{2},\widetilde{\a}\right) ,\\
	\mathcal{I}_m^{(\phi,s_1,s_2,\widetilde{\a})}
	&=\ds \frac{[N_m^{(\phi)}]^{2\widetilde{\a}}}{2^{\mu_x+\mu_y-1}} \mathcal{B}_{m}^{(\a,\b)}\left(\mu_y+2\widetilde{\a} \nu+\frac{1}{2},\mu_x+2\widetilde{\a} \xi+\frac{1}{2},\widetilde{\a}\right),
	\ea \right\}\widetilde{\a}\in\mathbb{N}.
	\eeq
	where 
	\beq
	\ba{ll}
	\mathcal{A}_{n}^{(\a)}(a,b,c)&=\ds  \int_0^{\infty} e^{-as} s^{b-1}\left[{}_1F_1\left(-n,\a,s\right)\right]^{2c}ds=\ds \sum_{i=0}^{2nc} \frac{(2c)! \widetilde{B}_{i+2c,2c} \left( c^{(n,\a)}_0,\cdots,(i+1)! c^{(n,\a)}_{i}\right)\Gamma\left(b + i\right)}{(i + 2c)!a^{b+i}},~c\in\mathbb{N},\\
	\mathcal{B}_{n}^{(\a,\b)}(a,b,c)&=\ds\int_{-1}^1 (1-x)^{a-1}(1+x)^{b-1} \left[P_n^{(\a,\b)}(x) \right]^{2c}dx\\
	&=\ds \left[\frac{(\a+1)_{n}}{n!}\right]^{2c}\sum\limits_{i=0}^{2cn}\frac{(2c)!\widetilde{B}_{2c+i,2c}\left(d_0^{(n,\a,\b)},\cdots,(i+1)!d_i^{(n,\a,\b)}\right)\mathcal{B}\left(a+i,b\right)}{(2c+i)!2^{1-a-b-i}} ,~c\in\mathbb{N},
	\ea
	\eeq 
	with
$c_i^{(n,\a)}= \frac{(-n)_i}{(\a)_i\,i!}$ (for $i\le n$), 0 (for $i>n$), whereas 
	$d_i^{(n,\a,\b)}=\frac{(-n)_i (n+\a+\b+1)_i}{(\a+1)_i 2^i i!}$ (for $i\le n$), $d_i^{(n,\a,\b)}=0$ (for $i>n$).
	
\begin{figure}[t]
	\centering
	\includegraphics[width=18cm,height=10cm]{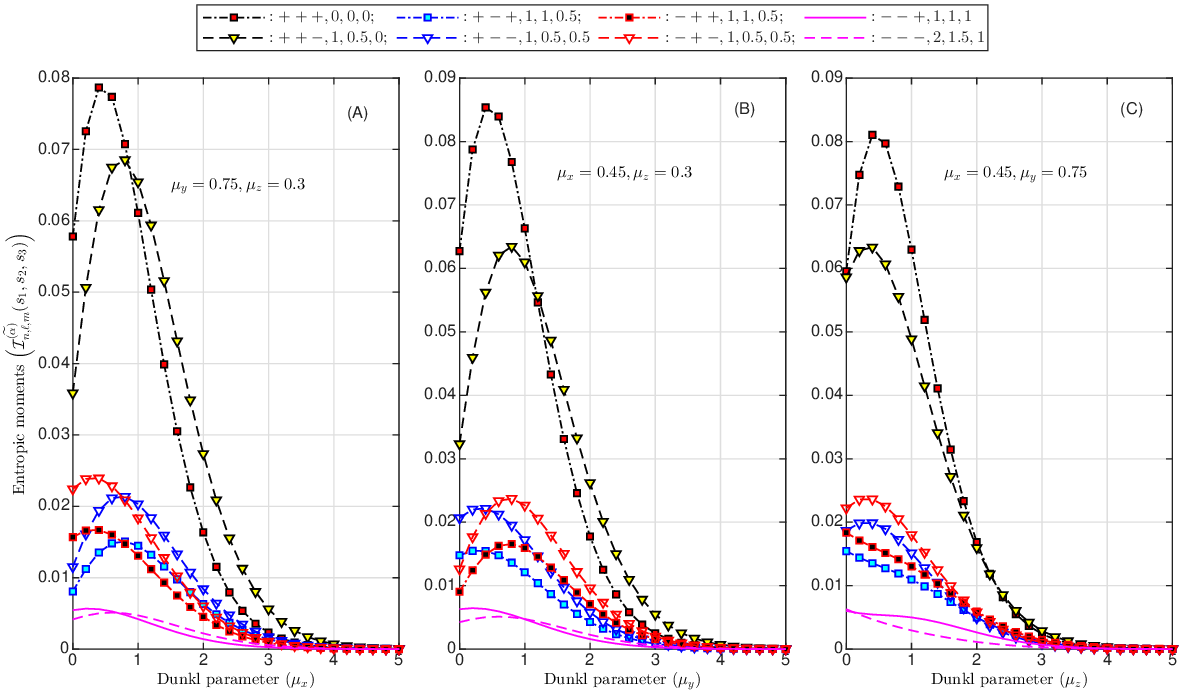}
	\caption{\label{fig2.entropic_moment}Comparison of entropic moments on total density function of order $\widetilde{\a}=2.25$ of different quantum states, having even and odd reflections. Reflection operators have eigenvalues $s_1,s_2,s_3$ and state quantum numbers $n,\ell,m$ are provided at the top. Fixed parameters are $\hbar=\mu=\omega=1$.}
\end{figure}

	To realize the effect of Dunkl parameters on information measures, in Fig.~\ref{fig2.entropic_moment}, we have plotted the entropic moments of total density function with respect to $\mu_x, \mu_y, \mu_z$ for $n=0,1,2$ for fixed $\ell=0,\frac{1}{2},1,\frac{3}{2}$ values, and $m=0,\frac{1}{2},1$, for eight parity sets: $s_1=\pm1, s_2=\pm1,s_3=\pm1$ with $\mu=\hbar=\omega=1$. The three panels, from left to right, correspond to the measures, with respect to (A) $\mu_x$, for fixed $\mu_y= 0.75, \mu_z=0.3$; (B) $\mu_y$ for fixed $\mu_x=-0.45, \mu_z=0.3$; (C) $\mu_z$, for fixed $\mu_x=-0.45, \mu_y=0.75$. The quantum numbers $n,\ell,m$ are produced along with the eight parities $+++$, $++-$, $+-+$, $+--$, $-++$, $-+-$, $--+$, $---$ in the legend. It is apparent that, in all cases, the moments traverse through a maximum value, before eventually falling down to zero as the parameters take progressively larger values. 

	Now, $\mathcal{R}$ of a density function $\rho_{n,\ell,m}^{(s_1,s_2,s_3)}$ is defined as \cite{renyi}: 
	\beq
	\mathcal{R}_{n,\ell,m}^{(s_1,s_2,s_3,\widetilde{\a})}=\frac{1}{1-\widetilde{\a}}\ln\left[\mathcal{I}_{n,\ell,m}^{(s_1,s_2,s_3,\widetilde{\a})}\right],\widetilde{\a}>0,\ne 1.
	\eeq 
    It satisfies the additive property for independent density functions and can be expressed as a sum of R\'enyi entropies of marginal density functions given below: 
	\beq
	\mathcal{R}_{n,\ell,m}^{(s_1,s_2,s_3,\widetilde{\a})}=\mathcal{R}_{n,\ell,m}^{(r,\widetilde{\a})}+\mathcal{R}_{\ell,m}^{(\theta,s_3,\widetilde{\a})}+\mathcal{R}_m^{(\phi,s_1,s_2,\widetilde{\a})},
	\eeq 
	where $\mathcal{R}_{n,\ell,m}^{(r,\widetilde{\a})} =\frac{1}{1-\widetilde{\a}}\ln\left[\mathcal{I}_{n,\ell,m}^{(r,\widetilde{\a})}\right]$, $\mathcal{R}_{\ell,m}^{(\theta,s_3,\widetilde{\a})}=\frac{1}{1-\widetilde{\a}}\ln\left[\mathcal{I}_{\ell,m}^{(\theta,s_3,\widetilde{\a})}\right]$ and $\mathcal{R}_m^{(\phi,s_1,s_2,\widetilde{\a})}=\frac{1}{1-\widetilde{\a}}\ln\left[\mathcal{I}_m^{(\phi,s_1,s_2,\widetilde{\a})}\right]$ for $\widetilde{\a}>0,\ne 1$.
	It has many applications in various quantum phenomena like communication protocol, disordered system, multifractal thermodynamics, entanglement, localization in phase space, molecular reactivity, quantum-classical correspondence, signal processing, chaos analysis for measure of complexity \cite{varga2003, renner2005, bialas2006} etc. For security analysis in quantum key distributions, it is used to measure entanglement and distinguishability of quantum states for $\widetilde{\a}\rightarrow\infty$. For $\widetilde{\a}=2$, it is used in machine learning and data clustering.
	
	Analogously, $\mathcal{T}$ of a density function $\rho_{n,\ell,m}^{(s_1,s_2,s_3)}$ can be written as \cite{tsallis},
	\beq
	\mathcal{T}_{n,\ell,m}^{(s_1,s_2,s_3,\widetilde{\a})}=-\ds\int[\rho_{n,\ell,m}^{(s_1,s_2,s_3)}(\mathbf{r})]^{(\widetilde{\a})}\ln_{\widetilde{\a}}\left[\rho_{n,\ell,m}^{(s_1,s_2,s_3)}(\mathbf{r})\right]d\chi,\widetilde{\a}>0,\ne 1,
	\eeq 
	where $\ln_{\widetilde{\a}}$ is the $\widetilde{\a}$-logarithmic function, given by $\ln_{\widetilde{\a}}(x)=\frac{x^{1-\widetilde{\a}}-1}{1-\widetilde{\a}}$, for $x>0$, $\widetilde{\a}\in\mathbb{R}-\{1\}$. In particular, when $\widetilde{\a}\rightarrow1$, $\ln_{\widetilde{\a}}(x)$ reduces to the natural logarithm function $\ln(x)$.
	It is a pseudo-additive measure of uncertainty having many applications in physics, statistical mechanics, non-equilibrium and chaotic systems, image processing 
	and so on. For $\widetilde{\a}<1$ and $\widetilde{\a} > 1$, the Tsallis entropy is sub-extensive and super-extensive respectively.
	Due to the pseudo-additivity nature, it can be derived as \cite{tsallis.pro}:
	\beq
	\ba{lr}
	\mathcal{T}^{(s_1,s_2,s_3,\widetilde{\a})}_{n,\ell,m}&=\ds\mathcal{T}^{(r,\widetilde{\a})}_{n,\ell,m}+\mathcal{T}^{(\theta,s_3,\widetilde{\a})}_{\ell,m}+\mathcal{T}^{(\phi,s_1,s_2,\widetilde{\a})}_{m}+(1-\widetilde{\a})\left[\mathcal{T}^{(r,\widetilde{\a})}_{n,\ell,m}\mathcal{T}^{(\theta,s_3,\widetilde{\a})}_{\ell,m}+\mathcal{T}^{(r,\widetilde{\a})}_{n,\ell,m}\mathcal{T}^{(\phi,s_1,s_2,\widetilde{\a})}_{m}+\mathcal{T}^{(\theta,s_3,\widetilde{\a})}_{\ell,m}\mathcal{T}^{(\phi,s_1,s_2,\widetilde{\a})}_{m}\right]\\
	&\quad  +(1-\widetilde{\a})^2\mathcal{T}^{(r,\widetilde{\a})}_{n,\ell,m}\mathcal{T}^{(\theta,s_3,\widetilde{\a})}_{\ell,m}\mathcal{T}^{(\phi,s_1,s_2,\widetilde{\a})}_{m}, 
	\ea 
	\eeq 
	where 
	$\mathcal{T}^{(r,\widetilde{\a})}_{n,\ell,m}=-\int_0^{\infty}\left[\rho_{n,\ell,m}^{(r)}\right]^{\widetilde{\a}}\ln_{\widetilde{\a}}\left[\rho_{n,\ell,m}^{(r)}\right] d\chi_{r}$, $\mathcal{T}^{(\theta,s_3,\widetilde{\a})}_{\ell,m}=-\int_0^{\pi}\left[\rho_{\ell,m}^{(\theta,s_3)}\right]^{\widetilde{\a}}\ln_{\widetilde{\a}}\left[\rho_{\ell,m}^{(\theta,s_3)}\right] d\chi_{\theta}$, and $\mathcal{T}^{(\phi,s_1,s_2,\widetilde{\a})}_{m}=-\int_0^{2\pi}\left[\rho_{m}^{(\phi,s_1,s_2)}\right]^{\widetilde{\a}}\ln_{\widetilde{\a}}\left[\rho_{m}^{(\phi,s_1,s_2)}\right] d\chi_{\phi}$, for $\widetilde{\a}>0,\ne 1$ are Tsallis entropies of the marginal density functions $\rho_{n,\ell,m}^{(r)}$, $\rho_{\ell,m}^{(\theta,s_3)}$ and $\rho_{m}^{(\phi,s_1,s_2)}$ respectively.
	It is clear from operators $N_{\theta}$ and $B_{\phi}$ that angular wave functions are directly reflected by reflection operators. To realize the effect of reflection on information measures, we have plotted $\mathcal{R}$ and $\mathcal{T}$ of radial wave function in Fig.~\ref{fig2.reflection.r} (B), (C) with respect to $n$; of angular $\theta$ function in Fig.~\ref{fig3.reflection.h} (B), (C) with respect to $\ell$; and of angular $\phi$ function in Fig.~\ref{fig4.reflection.g} (B), (C) with respect to $m$ quantum number respectively. 
	
	\subsubsection{Shannon entropy}\label{sec3.shannon}
	In this subsection, we will calculate $\mathcal{S}$ in $\xi\left(\mathbb{R}^*\times[0,\pi]\times[0,2\pi],d\chi\right)$ space. In a given state $\psi_{n,\ell,m}^{(s_1,s_2,s_3)}$, it is defined as, 
	\beq
	\ba{ll}
	\mathcal{S}_{n,\ell,m}^{(s_1,s_2,s_3)}&=-\ds\int\rho_{n,\ell,m}^{(s_1,s_2,s_3)}(\mathbf{r})\ln[\rho_{n,\ell,m}^{(s_1,s_2,s_3)}(\mathbf{r})]d\chi=\mathcal{S}_{n,\ell,m}^{(r)}+\mathcal{S}_{\ell,m}^{(\theta,s_3)}+\mathcal{S}_m^{(\phi,s_1,s_2)},
	\ea 
	\eeq 
	where $\mathcal{S}_{n,\ell,m}^{(r)}$, $\mathcal{S}_{\ell,m}^{(\theta,s_3)}$, $\mathcal{S}_{m}^{(\phi,s_1,s_2)}$ are entropies of marginal density functions. The radial part, $\mathcal{S}_{n,\ell,m}^{(r)}$ is derived as:  
	\beq
	\ba{ll}
	\mathcal{S}_{n,\ell,m}^{(r)}&=-\ds\int_0^{\infty} \rho_{n,\ell,m}^{(r)}\ln\left[\rho_{n,\ell,m}^{(r)}\right]d\chi_r=-\ln[(N_{n,\ell,m}^{(r)})^2]+\lam\left\langle r^2\right\rangle_{n,\ell,m}+\mathcal{S}_{n,\ell,m}^{(r,1)}+(2a-2L-2)\left\langle \ln[r]\right\rangle_{n,\ell,m},
	\ea
	\eeq
	where
	\beq
	\ba{ll}
	\left\langle \ln[r]\right\rangle_{n,\ell,m}&=\ds\frac{ [N_{n,\ell,m}^{(r)}]^2}{4\lam^{ L + \frac{3}{2}}}\sum_{i=0}^{2n} \frac{B_{i+2,2}\left( C^{(n)}_0, 2! C^{(n)}_1,..., (i+1)! C^{(n)}_i  \right)\Gamma \left( L + i + \frac{3}{2} \right) \Psi^{(d)} \left( L + i+ \frac{3}{2}  \right)}{(2 + i)!}-\frac{1}{2} \ln[\lam],
	\ea 
	\eeq
	and $\Psi^{(d)}(x)=\frac{1}{\G(x)}\frac{d\G(x)}{dx}$ denotes the digamma function, while 
	\beq
	\ba{ll}\label{sha.r1}
	\mathcal{S}_{n}^{(r,1)}(a)&=-\ds [N_{n,\ell,m}^{(r)}]^2\int_0^{\infty}e^{-\lam r^2}r^{2L+2}\left[{}_1F_1\left(-n;L + \frac{3}{2};\lam r^2\right)\right]^2\ln\left(\left[{}_2F_1\left(-n;L + \frac{3}{2};\lam r^2\right)\right]^2\right)dr\\
	&=-\ds\ln[(a_n^{(F)})^2]+\ds\frac{[N_{n,\ell,m}^{(r)}]^2}{2\lam^{L+\frac{3}{2}}}\sum\limits_{j=1}^{n}\left(\mathcal{C}^{(L + \frac{3}{2})}_{n}(L + \frac{3}{2},s_{n,j})+\mathcal{D}_{n}^{(L + \frac{3}{2})}(L + \frac{3}{2},s_{n,j})\right),
	\ea
	\eeq
	$0\le s_{n,1}\le s_{n,2}\le \cdots\le s_{n,n}<\infty$ are $n$ real roots of ${}_2F_1\left(-n;L+\frac{3}{2};s\right)=0$, $a_n^{(F)}=c_{n}^{(n,L+\frac{3}{2})}$ is the leading term of ${}_1F_1\left(-n;L + \frac{3}{2};s\right)$ and 
\beq 
\ba{ll}
\mathcal{C}_{n}^{(\a)}(a,x)&=-\ds\int_0^{x}e^{-s}s^{a-1}\left[{}_1F_1\left(-n;\a;s\right)\right]^2\ln\left(x-s\right)^2ds,\\	&=\ds-\ln[x^2]\mathcal{\bar{K}}_{n}^{(\a)}(a,x)-\sum\limits_{i=0}^{2n}\sum\limits_{k=0}^{\infty}\frac{(-1)^k 4x^{a+k+i}\widetilde{B}_{i+2,2}\left(c_0^{(n,\a)},\cdots,(i+1)!c_i^{(n,\a)}\right) \left[\Psi^{(d)}(1)-\Psi^{(d)}(a+k+i)\right]}{(i+2)!k!(a+k+i)},
\ea	
\eeq 
\beq
\ba{ll}
\mathcal{\bar{K}}_{n}^{(\a)}(a,x)&=\ds\int_0^{x}e^{-s}s^{a-1}\left[{}_1F_1\left(-n;\a;s\right)\right]^2ds=\ds\mathcal{A}_{n}^{(\a)}(1,a,1)-\sum\limits_{i=0}^{2n}\frac{2\widetilde{B}_{i+2,2}\left(c_0^{(n,\a)},\cdots,(i+1)!c_i^{(n,\a)}\right)\G(x,a+i)}{(i+2)!},
\ea
\eeq 
\beq 
\ba{ll}
\mathcal{D}_{n}^{(\a)}(a,x)&=-\ds\int_{x}^{\infty}e^{-s}s^{a-1}\left[{}_1F_1\left(-n;\a;s\right)\right]^2\ln\left(s-x\right)^2ds\\
&=-\ds\sum\limits_{i=0}^{2n}\sum \limits_{j=0}^{\lfloor a+i-1\rfloor}\frac{4\widetilde{B}_{i+2,2}\left(c_0^{(n,\a)},\cdots,(i+1)!c_i^{(n,\a)}\right)\, \binom{a+i-1}{j}e^{-x}\, x^{a+i-j-1} \, j!\, \Psi^{(d)}(j+1)}{(i+2)!}.
\ea
\eeq 
The convergent series $\sum \limits_{j=0}^{\lfloor a+i-1\rfloor}$ contains finite number of terms, when $a\in \mathbb{N}$; otherwise it is infinite.

\begin{figure}[t]
		\centering
		\includegraphics[width=18cm,height=12cm]{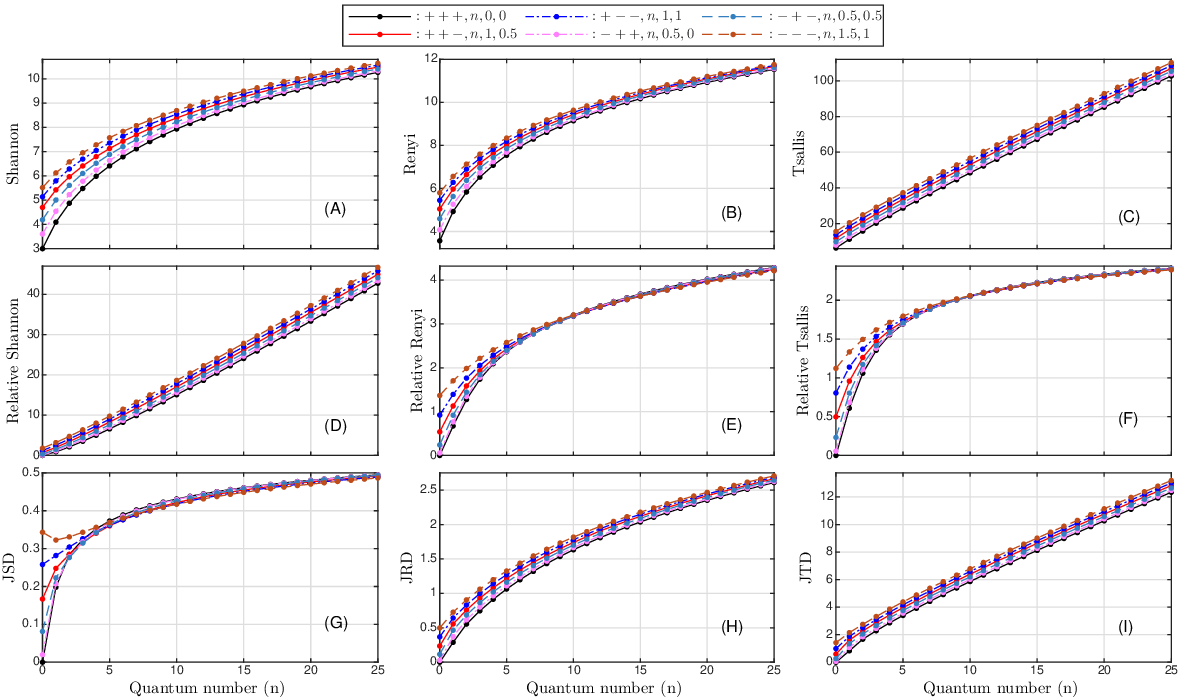}
			\caption{\label{fig2.reflection.r}The effect of $\widehat{R}_x, \widehat{R}_y, \widehat{R}_z$ on $\mathcal{S}$ of radial wave function, $R_{n,\ell,m}(r)/r^a$, w.r.t $n$ for 
			$\hbar=\mu=\omega=1$, $\mu_x=-0.45$, $\mu_y=0.75$, $\mu_z=0.3$. The order $\widetilde{\a}=0.7$ is used for (B), (C), (E), (F), (H) and (I).}
	    \end{figure}

	The Shannon entropy $\mathcal{S}^{(\theta,s_3)}_{\ell,m}$ of the angular density function $\rho_{\ell,m}^{(\theta,s_3)}(\theta)$ is found to be as follows:   
	\beq
	\mathcal{S}^{(\theta,s_3)}_{\ell,m}=-\ds\int_{0}^{\pi}\rho_{\ell,m}^{(\theta,s_3)}\ln[\rho_{\ell,m}^{(\theta,s_3)}]d\chi_{\theta}=-2\ln[N_{\ell,m}^{(\theta)}]+\mathcal{S}_{\ell,m}^{(\theta,s_3,1)}+2\sigma\mathcal{S}^{(\theta,s_3,2)}_{\ell,m}+2\varsigma\mathcal{S}^{(\theta,s_3,3)}_{\ell,m},
	\eeq
	where
	\beq\label{sha.theta1}
	\ba{ll}
	\mathcal{S}^{(\theta,s_3,1)}_{\ell,m}&=\ds-\frac{[N_{\ell,m}^{(\theta)}]^2}{2^{a-\frac{1}{2}}}\ds\int_{-1}^{1}(1-q)^{\g}(1+q)^{\d}\left[P_{\ell}^{(\g,\d)}(q)\right]^2\ln\left(\left[P_{\ell}^{(\g,\d)}(q)\right]^2\right)dq\\
	&=\ds-\ln\left[a_{\ell}^{(\g,\d)}\right]^2-\frac{[N_{\ell,m}^{(\theta)}]^2}{2^{a-\frac{1}{2}}}\sum\limits_{j=1}^{\ell}\left[\mathcal{E}^{(\g,\d)}_{\ell}(\g+1,\d+1,q_{\ell,j})+\mathcal{F}^{(\g,\d)}_{\ell}(\g+1,\d+1,q_{\ell,j})\right].
	\ea
	\eeq 
	The improper integral $\mathcal{E}^{(\a,\b)}_{\ell}(a,b,x)$ is given by, 
	\beq
	\ba{ll}
	\ds\mathcal{E}^{(\a,\b)}_{\ell}(a,b,x)&=\ds\int_{-1}^{x}(1-q)^{a-1}(1+q)^{b-1}\left[P_{\ell}^{(\a,\b)}(q)\right]^2\ln\left(x-q\right)^2dq\\
	&=\ds\frac{[(\a+1)_{\ell}]^2}{[\ell!]^2}\sum\limits_{i=0}^{2\ell}\sum\limits_{k=0}^{\lfloor a+i-1\rfloor}\left[\frac{4(1-x)^{a+i-k-1}(1+x)^{b+k}\binom{a+i-1}{k}\widetilde{B}_{i+2,2}\left(d_0^{(\ell,\a,\b)},\cdots,(i+1)!d_i^{(\ell,\a,\b)}\right)}{(i+2)!}\right]\\
	&\quad\times\mathcal{B}(k+1,b)\left[\Psi^{(d)}(k+1)-\Psi^{(d)}(b+k+1)\right]+\ln(1+x)^2\mathcal{J}_{\ell}^{(\a,\b)}(a,b,x),\\
	\ea
	\eeq
	whereas the improper integral $\mathcal{F}^{(\a,\b)}_{\ell}(a,b,x)$ is given by, 
	\beq 
	\ba{ll}
	\ds\mathcal{F}^{(\a,\b)}_{\ell}(a,b,x)&=\ds\int_{x}^{1}(1-q)^{a-1}(1+q)^{b-1}\left[P_{\ell}^{(\a,\b)}(q)\right]^2\ln\left(q-x\right)^2dq,\\
	&=\ds\frac{4[(\g+1)_{\ell}]^2}{[\ell!]^2}\sum\limits_{i=0}^{2\ell}\sum\limits_{k=0}^{\lfloor b-1\rfloor}\left[\frac{(1-x)^{a+i+k}(1+x)^{b-k-1}\binom{b-1}{k}\widetilde{B}_{i+2,2}\left(d_0^{(\ell,\a,\b)},\cdots,(i+1)!d_i^{(\ell,\a,\b)}\right)}{(i+2)!}\right]\\
	&\quad\times\mathcal{B}(k+1,a+i)\left[\Psi^{(d)}(k+1)-\Psi^{(d)}(a+k+i+1)\right]+\ln(1-x)^2\left[\mathcal{B}_{\ell}^{(\a,\b)}(a,b,1)-\mathcal{J}_{\ell}^{(\a,\b)}(a,b,x)\right], 
	\ea
	\eeq
	\beq 
	\ba{ll}
	\mathcal{J}_{n}^{(\a,\b)}(a,b,x)=\ds\int_{-1}^{x}(1-q)^{a-1}(1+q)^{b-1}\left[P_{n}^{(\a,\b)}(q)\right]^2dq\\
	=\ds\frac{[(\a+1)_{n}]^2}{[n!]^2}\sum\limits_{i=0}^{2n}\sum\limits_{k=0}^{\lfloor a+i-1\rfloor}\left[\frac{2(1-x)^{a+i-k-1}(1+x)^{b+k+1}\binom{a+i-1}{k}\widetilde{B}_{i+2,2}\left(d_0^{(n,\a,\b)},\cdots,(i+1)!d_i^{(n,\a,\b)}\right)}{(i+2)!}\right]\mathcal{B}(k+1,b), 
	\ea
	\eeq
	where $-1 \leq q_{\ell,1} < q_{\ell,2} < \cdots < q_{\ell,\ell} \leq 1$ are roots of $P_\ell^{(\g,\d)}(q) = 0$, $a_\ell^{(\g,\d)}=\frac{(\g+1)_{\ell} d_{\ell}^{(\ell,\g,\d)}}{\ell!}$ is the leading term of $P_\ell^{(\g,\d)}(q)$. Furthermore, 
	\beq
	\ba{ll}
	\mathcal{S}^{(\theta,s_3,2)}_{\ell,m}	=\ds-\frac{[N_{\ell,m}^{(\theta)}]^2}{2^{a-\frac{1}{2}}}\mathcal{K}_{\ell}^{(\g,\d)}(\g,\d),~
	\mathcal{S}^{(\theta,s_3,3)}_{\ell,m}	=\ds-\frac{[N_{\ell,m}^{(\theta)}]^2}{2^{a-\frac{1}{2}}}\mathcal{L}_{\ell}^{(\g,\d)}(\g,\d),
	\ea 
	\eeq
\beq 
\ba{ll}
\mathcal{K}_{n}^{(\a,\b)}(a,b)=\ds\int_{-1}^{1}(1-q)^{a-1}(1+q)^{b-1}\left[P_{n}^{(\a,\b)}(q)\right]^2\ln\left[1-q\right]dq\\
=\ds \left[\frac{(\a+1)_{n}}{n!}\right]^{2}\sum\limits_{i=0}^{2n}\frac{\widetilde{B}_{2+i,2}\left(d_0^{(n,\a,\b)},\cdots,(i+1)!d_i^{(n,\a,\b)}\right)2^{a+b+i} \mathcal{B}\left(a+i,b\right)}{(2+i)!}\left[\ln[2]+\Psi^{(d)}(a+i)-\Psi^{(d)}(a+b+i)\right],			
\ea		
\eeq 
\beq 
\ba{ll}		\mathcal{L}_{n}^{(\a,\b)}(a,b)=\ds\int_{-1}^{1}(1-q)^{a-1}(1+q)^{b-1}\left[P_{n}^{(\a,\b)}(q)\right]^2\ln\left[1+q\right]dq\\
=\ds \left[\frac{(\a+1)_{n}}{n!}\right]^{2}\sum\limits_{i=0}^{2n}\frac{\widetilde{B}_{2+i,2}\left(d_0^{(n,\a,\b)},\cdots,(i+1)!d_i^{(n,\a,\b)}\right)2^{a+b+i} \mathcal{B}\left(a+i,b\right)}{(2+i)!}\left[\ln[2]+\Psi^{(d)}(b)-\Psi^{(d)}(a+b+i)\right].
\ea
\eeq 
Similarly, the Shannon entropy $\mathcal{S}^{(\phi,s_1,s_2)}_m$ of the density function $\rho_{m}^{(\phi,s_1,s_2)}$ can be expressed as, 
	\beq
	\mathcal{S}^{(\phi,s_1,s_2)}_m=-\ds\int_{0}^{2\pi}\rho_{m}^{(\phi,s_1,s_2)}\ln[\rho_{m}^{(\phi,s_1,s_2)}(\phi)]d\chi_{\phi}=-\ln[(N_{m}^{(\phi)})^2]+\mathcal{S}^{(\phi,s_1,s_2,1)}_m+2\nu\mathcal{S}^{(\phi,s_1,s_2,2)}_m+2\xi\mathcal{S}^{(\phi,s_1,s_2,3)}_m, 
	\eeq
	where
	\beq\label{sh.phi1}
	\ba{ll}
	\mathcal{S}^{(\phi,s_1,s_2,1)}_m&=\ds-\frac{[N_{m}^{(\phi)}]^2}{2^{\mu_x+\mu_y-1}}\ds\int_{-1}^{1}(1-p)^{\a}(1+p)^{\b}\left[P_{m}^{(\a,\b)}(p)\right]^2\ln\left(\left[P_{m}^{(\a,\b)}(q)\right]^2\right)dp\\
	&=-\ds\frac{[N_{m}^{(\phi)}]^2}{2^{\mu_x+\mu_y-1}}\sum\limits_{j=1}^{m}\left[\mathcal{E}^{(\a,\b)}_{m}(\a+1,\b+1,p_{m,j})+\mathcal{F}^{(\a,\b)}_{m}(\a+1,\b+1,p_{m,j})\right]-\ln\left[a_{m}^{(\a,\b)}\right]^2,
	\ea
	\eeq 
where $-1 \leq p_1 < p_2 < \cdots < p_m \leq 1$ are the roots of $P_m^{(\a,\b)}(p) = 0$, and $a_m^{(\a,\b)}=\frac{(\a+1)_{m} d_{m}^{(m,\a,\b)}}{m!}$ is the leading term of $P_m^{(\a,\b)}(p)$. In a similar fashion, the other two terms can be derived as: 
	\beq
	\ba{ll}
	\mathcal{S}^{(\phi,s_1,s_2,2)}_m	=-\ds\frac{[N_{m}^{(\phi)}]^2}{2^{\mu_x+\mu_y-1}}\mathcal{K}_{m}^{(\a,\b)}(\a+1,\b+1) ,~
	\mathcal{S}^{(\phi,s_1,s_2,3)}_m	=-\ds\frac{[N_{m}^{(\phi)}]^2}{2^{\mu_x+\mu_y-1}}\mathcal{L}_{m}^{(\a,\b)}(\a+1,\b+1) .
	\ea
	\eeq
	The calculated marginal Shannon entropies are portrayed in top left (A) panels of Figs.~\ref{fig2.reflection.r}, \ref{fig3.reflection.h}, \ref{fig4.reflection.g} w.r.t. $n,\ell,m$ quantum numbers for the parameters listed in respective figures. It is seen that for $s_1=1,s_2=-1$ and $s_1=-1, s_2=1$ the entropies of radial wave functions represent the same results. Similarly, due to an indirect effect of $\widehat{R}_x, \widehat{R}_y$ on $H$, the marginal entropies for $s_1=1,s_2=-1$ and $s_1=-1, s_2=1$ represent same results but total entropies of overall function for any pair of distinct states represent different results. For a better appreciation, we plot $\mathcal{S}, \mathcal{R}, \mathcal{T}$ in top, middle, bottom panels of Fig.~\ref{fig5.reflection.total} with respect to $m$ for eight parities $+++$, $++-$, $+-+$, $+--$, $-++$, $-+-$, $--+$, $---$. Recently, the entropy of hydrogen-like ions in DSE for these parities were presented in \cite{arxiv.dn2}. In passing, it may be noted that, finding analytical results of Shannon entropy is an open problem for classical polynomials, as reflected from a lot of publications \cite{pra1985,jsd1994,log.pot,jsd2010,jsd2010.jacobi,jsd2011}. It is possible to extend this approach in momentum space \cite{chung2023}. In our previous study, we have provided entropies for Bessel functions \cite{dn.jmc2023} as well as hyper-geometric and Jacobi polynomials in DS system \cite{arxiv.dn}. The quasi-analytical expression in a confined harmonic oscillator in presence of a time‑dependent moving boundary has been reported in \cite{shannon.dn}.
	
	\subsection{Relative information}\label{sec4.relative}
	In this section, results are presented for various relative information measures of the isotropic harmonic oscillator. 
	\subsubsection{Relative entropy and generalized Jensen-Shannon divergence (GJSD)}
\begin{figure}[t]
	\centering		
	\includegraphics[width=18cm,height=12cm]{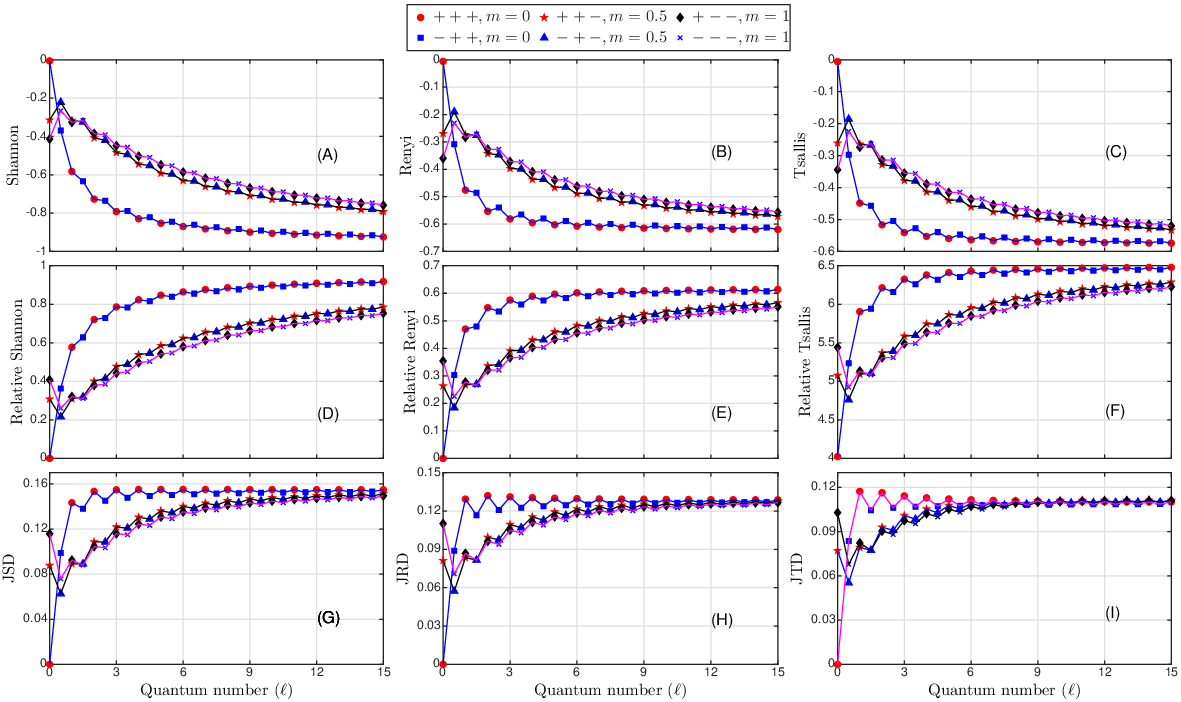}
	\caption{\label{fig3.reflection.h} Direct effect of $\widehat{R}_z$ and indirect effect of $\widehat{R}_x,\widehat{R}_y$ on various measures for $H_{\ell,m}^{(s_3)}(\theta)$ w.r.t. $\ell$ for $\hbar=\mu=\omega=1$, $m=0$, $\mu_x=0.45$, $\mu_y=0.4$, $\mu_z=0.3$.  The order $\widetilde{\a}=0.75$ is used for (B), (C), (E), (F), (H) and (I).}
\end{figure}

	The Kullback-Leibler entropy or relative entropy is a non-negative function, which provides a measure of how one probability distribution
	differs from the another, and as such, defined by \cite{kb,kb2,relative.shannon}, 
	\beq
	\ba{ll}
	\ds S_{KL}(\rho_{n,\ell,m}^{(s_1,s_2,s_3)}(\mathbf{r}),\rho_{n',\ell',m'}^{(s_1',s_2',s_3')}(\mathbf{r}))=\ds\int\rho_{n,\ell,m}^{(s_1,s_2,s_3)}(\mathbf{r})\ln\left(\f{\rho_{n,\ell,m}^{(s_1,s_2,s_3)}(\mathbf{r})}{\rho_{n',\ell',m'}^{(s_1',s_2',s_3')}(\mathbf{r})}\right)d\chi.
	\ea 
	\eeq
    Further simplification leads to, 
	\beq
	\ba{ll}
	\ds S_{KL}(\rho_{n,\ell,m}^{(s_1,s_2,s_3)}(\mathbf{r}),\rho_{n',\ell',m'}^{(s_1',s_2',s_3')}(\mathbf{r}))&=S_{KL}\left(\rho_{n,\ell,m}^{(r)},\rho_{n',\ell',m'}^{(r)}\right)+S_{KL}\left(\rho_{\ell,m}^{(\theta,s_3)},\rho_{\ell',m'}^{(\theta,s_3')}\right)+S_{KL}\left(\rho_{m}^{(\phi,s_1,s_2)},\rho_{m'}^{(\phi,s_1',s_2')}\right),
	\ea 
	\eeq
	where the relative entropies of marginal radial $r$, as well as angular $\theta$ and $\phi$ density functions are expressed as,   
	\beq
	\ba{lr}
	\ds S_{KL}\left(\rho_{n,\ell,m}^{(r)},\rho_{n',\ell',m'}\right)
	=\ds\ln\frac{[N_{n,\ell,m}^{(r)}]^2}{[N_{n',\ell',m'}^{(r)}]^2}
	+\left\langle\ln\left(\frac{{}_1F_1\left(-n,L+\frac{3}{2},\lam r^2\right)}{{}_1F_1\left(-n',L'+\frac{3}{2},\lam r^2\right)}\right)^2\right\rangle_{n,\ell,m}+2(L-L')\langle \ln[r]\rangle_{n,\ell,m},\\
	\ds S_{KL}\left(\rho_{\ell,m}^{(\theta,s_3)},\rho_{\ell',m'}^{(\theta,s_3')}\right)
	=\ds\ln\frac{[N_{\ell,m}^{(\theta)}]^2}{[N_{\ell',m'}^{(\theta)}]^2}+\left\langle\ln\left[\frac{[P_{\ell}^{(\g,\d)}(\cos 2\theta)]^2}{[P_{\ell'}^{(\g,\d)}(\cos 2 \theta)]^2}\right]\right\rangle_{\ell,m}+2(\sigma-\sigma')\mathcal{S}^{(\theta,s_3,2)}_{\ell,m}+2(\varsigma-\varsigma')\mathcal{S}^{(\theta,s_3,3)}_{\ell,m},\\
   S_{KL}\left(\rho_{m}^{(\phi,s_1,s_2)},\rho_{m'}^{(\phi,s_1',s_2')}\right)
	=\ds\ln\frac{[N_{m}^{(\phi)}]^2}{[N_{m'}^{(\phi)}]^2}+\left\langle\ln\left[\frac{[P_{m}^{(\a,\b)}(\cos 2 \phi)]^2}{[P_{m'}^{(\a,\b)}(\cos 2 \phi)]^2}\right]\right\rangle_m+2(\nu-\nu')\mathcal{S}_{m}^{(\phi,s_1,s_2,2)}+2(\xi-\xi')\mathcal{S}_{m}^{(\phi,s_1,s_2,3)},
	\ea
	\eeq
	with $L'=-\frac{1}{2}+\sqrt{a(a-1)-\eps^{(2)}_{\ell',m'}}$, and $\sigma'$, $\varsigma'$, $\nu'$, $\xi'$ are obtained from the equation (\ref{sigma.varsigma}) for $\ell=\ell'$, $m=m'$, $s_1=s_1'$, $s_2=s_2'$, $s_3=s_3'$. It may be noted that, 
	$S_{KL}\left(\rho,\rho^{'}\right)$ is non-negative, and it represents the average information gain of the density $\rho$ for a given density $\rho'$ satisfying the 
	asymmetric condition $S_{KL}\left(\rho,\rho^{'}\right)\ne S_{KL}\left(\rho',\rho\right)$ and $S_{KL}\left(\rho,\rho^{'}\right)=0$ if and only if $\rho=\rho'$.
	These relative entropies of marginal density functions take particularly simpler form for the special case: $n'=\ell'=m'=0$, with parities $+++$. These are given as below,  
	\beq
	\ba{ll}
	\ds S_{KL}\left(\rho_{n,\ell,m}^{(r)},\rho_{0,0,0}^{(r)}\right)=\ds\ln\frac{[N_{n,\ell,m}^{(r)}]^2}{[N_{0,0,0}^{(r)}]^2}+\mathcal{S}_{n,\ell,m}^{(r,1)}+2(L-L_{0})\left\langle \ln[r]\right\rangle_{n,\ell,m}^{(r)},\\
	\ds S_{KL}\left(\rho_{\ell,m}^{(\theta,s_3)},\rho_{0,0}^{(\theta,s_3')}\right)=\ds\ln\frac{[N_{\ell,m}^{(\theta)}]^2}{[N_{0,0}^{(\theta)}]^2}+ \mathcal{S}^{(\theta,s_3,1)}_{\ell,m}+2(\sigma-\sigma_0) \mathcal{S}^{(\theta,s_3,2)}_{\ell,m}+2(\varsigma-\varsigma_0)\mathcal{S}_{\ell,m}^{(\theta,s_3,3)},\\
	\ds S_{KL}\left(\rho_{m}^{(\phi,s_1,s_2)},\rho_{0}^{(\phi,s_1',s_2')}\right)=\ds\ln\frac{[N_{m}^{(\phi)}]^2}{[N_{0}^{(\theta)}]^2}+ \mathcal{S}_{m}^{(\phi,s_1,s_2,1)}+2(\nu-\nu_0)\mathcal{S}_m^{(\phi,s_1,s_2,2)}+2(\xi-\xi_0)\mathcal{S}_{m}^{(\phi,s_1,s_2,3)}
	\ea
	\eeq
	where $L_0=-\frac{1}{2}+\sqrt{a(a-1)-\eps^{(2)}_{0,0}}$, $\sigma_0=\varsigma_0=\nu_0=\xi_0=0$.  
	The calculated relative Shannon entropy of marginal densities, $\rho_{n,\ell,m}^{(r)}$,  $\rho_{\ell,m}^{(\theta,s_3)}$, $\rho_{m}^{(\phi,s_1,s_2)}$, 
	are portrayed in middle left (D) panels of Figs.~\ref{fig2.reflection.r}, \ref{fig3.reflection.h}, \ref{fig4.reflection.g} w.r.t. $n,\ell,m$ quantum numbers, for set of parameters 
	listed in respective figures. 
	
	Now, suppose $\{\rho_1,\rho_2,\cdots,\rho_n\}$ is a sequence of density functions, and $\{\lambda_1,\lambda_2,\cdots,\lambda_n\}$ is a sequence of weights such that $\sum_{i=1}^{n}\lambda_i=1$. Then the generalized Jensen-Shannon divergence (GJSD) is defined by \cite{gjsd},
	\beq
	GJSD(\rho_1,\rho_2,\cdots,\rho_n;\lambda_1,\lambda_2,\cdots,\lambda_n)=\mathcal{S}\left(\sum_{i=1}^{n}\lambda_i\rho_i\right)-\sum_{i=1}^{n}\lambda_i\mathcal{S}\left(\rho_i\right). 
	\eeq 
	It satisfies several properties \cite{gjsd,gjsd.pro,gjsd.pro2,gjsd.pro3}, having many applications in biology and cluster analysis \cite{gjsd.appl}, distance measure between random graphs, time series \cite{gjsd.appl3}, statistical analysis \cite{gjsd.appl4} etc. The GJSD between two states of a DS system, having densities, $\rho_{n,\ell,m}^{(s_1,s_2,s_3)} (\mathbf{r}), \rho_{n',\ell',m'}^{(s_1',s_2',s_3')}(\mathbf{r})$, with weights $\upsilon$ and $(1-\upsilon)$, can be written \cite{jsd,jsd2} as, 
	\beq
	\ba{lr}
	\ds GJSD(\rho_{n,\ell,m}^{(s_1,s_2,s_3)}(\mathbf{r}),\rho_{n',\ell',m'}^{(s_1',s_2',s_3')}(\mathbf{r});\upsilon,1-\upsilon)&=\mathcal{S}\left(\upsilon\rho_{n,\ell,m}^{(s_1,s_2,s_3)}(\mathbf{r})+(1-\upsilon)\rho_{n',\ell',m'}^{(s_1',s_2',s_3')}(\mathbf{r})\right)-\upsilon\mathcal{S}(\rho_{n,\ell,m}^{(s_1,s_2,s_3)}(\mathbf{r}))\\
	&-(1-\upsilon)\mathcal{S}(\rho_{n',\ell',m'}^{(s_1',s_2',s_3')}(\mathbf{r})).
	\ea 
	\eeq
	For $\upsilon=\frac{1}{2}$, we denote JSD by  $JSD\left(\rho_{n,\ell,m}^{(s_1,s_2,s_3)}(\mathbf{r}),\rho_{n',\ell',m'}^{(s_1',s_2',s_3')}(\mathbf{r})\right)$, which can be written as,   
	\beq
	\ba{lr}
	JSD \left(\rho_{n,\ell,m}^{(s_1,s_2,s_3)}(\mathbf{r}),\rho_{n',\ell',m'}^{(s_1',s_2',s_3')}(\mathbf{r}) \right)&=\ds\sum\limits_{i=1}^{\infty}\frac{(-1)^{i}}{2i} \left[\mathcal{J}^{(r,\theta,\phi)}_{(n,\ell,m),(n',\ell',m')}(i+1)+\mathcal{J}^{(r,\theta,\phi)}_{(n,\ell,m),(n',\ell',m')}(i)\right]+\varXi^{(r)}+\varXi^{(\theta)}+\varXi^{(\phi)}\\
	&\ds-\ln[4]-\frac{1}{2}\left[\mathcal{S}(\rho_{n,\ell,m}^{(s_1,s_2,s_3)}(\mathbf{r}))+\mathcal{S}(\rho_{n',\ell',m'}^{(s_1',s_2',s_3')}(\mathbf{r}))\right],
	\ea
	\eeq 
	where 
	\beq
	\ba{ll}
	\mathcal{J}^{(r,\theta,\phi)}_{(n,\ell,m),(n',\ell',m')}(\widetilde{\a})&=\mathcal{J}_{n,n'}^{(r)}(\widetilde{\a})\mathcal{J}_{\ell,\ell'}^{(\theta)}(\widetilde{\a})\mathcal{J}_{m,m'}^{(\phi)}(\widetilde{\a}),
	\ea
	\eeq 
	signifies the relative entropic moment between the joint density functions, $\rho_{n,\ell,m}^{(s_1,s_2,s_3)}(\mathbf{r})$ and $\rho_{n',\ell',m'}^{(s_1',s_2',s_3')}(\mathbf{r})$ of order $\widetilde{\a}$. The latter measure, for marginal densities, are found as, 
	$\ds\mathcal{J}_{(n,\ell,m),(n',\ell',m')}^{(r)}(\widetilde{\a})=\int_0^{\infty}\frac{[\rho_{n,\ell,m}^{(r)}]^{\widetilde{\a}}}{[\rho_{n',\ell',m'}^{(r)}]^{{\widetilde{\a}}-1}}d\chi_{r}$, $\ds\mathcal{J}_{{(\ell,m)},{(\ell',m)}}^{(\theta)}(\widetilde{\a})=\int_0^{\pi}\frac{[\rho_{\ell,m}^{(\theta,s_3)}]^{\widetilde{\a}}}{[\rho_{\ell',m'}^{(\theta,s'_3)}]^{{\widetilde{\a}}-1}}d\chi_{\theta}$ and $\ds\mathcal{J}_{m,m'}^{(\phi)}(\widetilde{\a})=\ds\int_0^{2\pi}\frac{[\rho_{m}^{(\phi,s_1,s_2)}]^{\widetilde{\a}}}{[\rho_{m'}^{(\phi,s'_1,s'_2)}]^{{\widetilde{\a}}-1}}d\chi_{\phi}$.
	The terms $\varXi^{(r)}, \varXi^{(\theta)}, \varXi^{(\phi)}$ are given as below,
	\beq
	\ba{ll}
	\varXi^{(r)}&=\ds \ln 2- 2\ln [N_{n',\ell',m'}^{(r)}] + \frac{\lam\left[ \langle r^2 \rangle_{n,\ell,m}^{(r)} + \langle r^2 \rangle_{n',\ell',m'}^{(r)} \right]}{2}  -(L'+1-a)  \left[ \langle \ln r \rangle_{n,\ell,m}^{(r)} + \langle \ln r \rangle_{n',\ell',m'}^{(r)} \right],\\
	\varXi^{(\theta)}&=\ds \ln 2\ds-2\ln[N_{\ell',m'}^{\theta)}]-\sigma'\left[\mathcal{S}^{(\theta,s_3,2)}_{\ell,m} + \mathcal{S}^{(\theta,s_3',2)}_{\ell',m'}\right]
	-\varsigma'\left[\mathcal{S}^{(\theta,s_3,3)}_{\ell,m} + \mathcal{S}^{(\theta,s_3',3)}_{\ell',m'}\right],\\
	\varXi^{(\phi)}&=\ds \ln 2 \ds-2 \ln[N_{m'}^{(\phi)}]-\nu'\left[\mathcal{S}^{(\phi,s_1,s_2,2)}_m + \mathcal{S}^{(\phi,s_1',s_2',2)}_{m'}\right]-\xi'\left[\mathcal{S}^{(\phi,s_1,s_2,3)}_m+\mathcal{S}^{(\phi,s_1',s_2',3)}_{m'}\right].
	\ea
	\eeq 
	Note that $JSD(\rho,\rho')\ge0$, finite, fulfils the property $JSD(\rho,\rho')=JSD(\rho',\rho)$, and $JSD(\rho,\rho')=0$ if and only if $\rho=\rho'$. The square root 
	$\sqrt{JSD(\rho,\rho')}$ satisfies the inequality $\sqrt{JSD(\rho,\rho')} + \sqrt{JSD(\rho',\rho^{''})} \ge \sqrt{JSD(\rho,\rho^{''})}$. 
	Now, the JSD of marginal density functions are given by: 
	\beq 
	\ba{ll}
	JSD\left(\rho_{n,\ell,m}^{(r)},\rho_{n',\ell',m'}^{(r)}\right)
	&= \varXi^{(r)} + \ds\sum\limits_{i=1}^{\infty}\frac{(-1)^{i}}{2i} \left[\mathcal{J}_{n,n'}^{(r)}(i+1)+\mathcal{J}_{n,n'}^{(r)}(i)\right]-\frac{1}{2}\left(\mathcal{S}_{n,\ell,m}^{(r)}+\mathcal{S}_{n',\ell',m'}^{(r)}\right),  
	\ea
	\eeq
	\beq 
	\ba{ll}
	JSD\left(\rho_{\ell,m}^{(\theta,s_3)},\rho_{\ell',m'}^{(\theta,s_3')}\right)
	&=\varXi^{(\theta)}+\ds\sum\limits_{i=1}^{\infty}\frac{(-1)^{i}}{2i} \left[\mathcal{J}_{\ell,\ell'}^{(\theta)}(i+1)+\mathcal{J}_{\ell,\ell'}^{(\theta)}(i)\right]-\frac{1}{2}\left(\mathcal{S}_{\ell,m}^{(\theta,s_3)}+\mathcal{S}_{\ell',m'}^{(\theta,s_3')}\right),
	\ea
	\eeq
	\beq 
	\ba{ll}JSD\left(\rho_{m}^{(\phi,s_1,s_2)},\rho_{m'}^{(\phi,s_1',s_2')}\right)
	&=\ds\varXi^{(\phi)}+\sum\limits_{i=1}^{\infty}\frac{(-1)^{i}}{2i} \left[\mathcal{J}_{m,m'}^{(\phi)}(i+1)+\mathcal{J}_{m,m'}^{(\phi)}(i)\right]-\frac{1}{2}\left(\mathcal{S}_{m}^{(\phi,s_1,s_2)}+\mathcal{S}_{m'}^{(\phi,s_1',s_2')}\right).
	\ea
	\eeq
	The relative entropic moments between two arbitrary states cannot be obtained analytically for a given order. But one can define $ \mathcal{J}^{(r,\theta,\phi)}_{(n,\ell,m),(0,0,0)}(i)$ for (+)ve integral order $i\in\mathbb{N}$. For marginal density functions, they are derived as: 
	\beq
	\ba{ll}
		\mathcal{J}_{n,0}^{(r)}(i)&=\ds\frac{(N_{n,\ell,m}^{(r)})^{2i}\mathcal{A}_{n}^{(L+\frac{3}{2})}\left(1, i(L-L')+L'+\frac{1}{2} ,i\right)}{2(N_{0,0,0}^{(r)})^{2i-2}\lam^{i(L-L')+L'+\frac{3}{2}}},\\
			\mathcal{J}_{\ell,0}^{(\theta)}(i)&=\ds\frac{(N_{\ell,m}^{(\theta)})^{2i}}{(N_{0,0}^{(\theta)})^{2i-2}2^{a-\frac{1}{2}}}\mathcal{B}_{\ell}^{(\g,\d)}\left(\mu_x+\mu_y+2i(\sigma-\sigma')i+2\sigma'+1,\mu_z+2i(\varsigma-\varsigma')+2\varsigma'+\frac{1}{2},i\right),\\
			\mathcal{J}_{m,0}^{(\phi)}(i)&=\ds\frac{(N_{m}^{(\phi)})^{2i}}{(N_{0}^{(\phi)})^{2i-2}2^{\mu_x+\mu_y-1}}\mathcal{B}_{m}^{(\a,\b)}\left(2i(\nu-\nu')+2\nu'+\mu_y+\frac{1}{2},2i(\xi-\xi')+2\xi'+\mu_x+\frac{1}{2},i\right).
		\ea
	\eeq
	It may be mentioned that, JSD is non-negative, and $JSD(\rho_{n,\ell,m}^{(s_1,s_2,s_3)}(\mathbf{r}),\rho_{n',\ell',m'}^{(s_1',s_2',s_3')}(\mathbf{r}))=0$, if and only if $\rho_{n,\ell,m}^{(s_1,s_2,s_3)}(\mathbf{r})=\rho_{n',\ell',m'}^{(s_1',s_2',s_3')}(\mathbf{r})$. The calculated JSD of $\rho_{n,\ell,m}^{(r)}$,  $\rho_{\ell,m}^{(\theta,s_3)}$ and $\rho_{m}^{(\phi,s_1,s_2)}$ , 
	are displayed in bottom left (G) panels of Figs.~\ref{fig2.reflection.r}, \ref{fig3.reflection.h}, \ref{fig4.reflection.g}, w.r.t. $n, \ell, m$ quantum numbers, corresponding to the 
	parameters, which are given in the figures.

\subsubsection{Relative R\'enyi and generalized Jensen-R\'enyi divergence (GJRD)}
\begin{figure}[t]
	\centering
	\includegraphics[width=18cm,height=12cm]{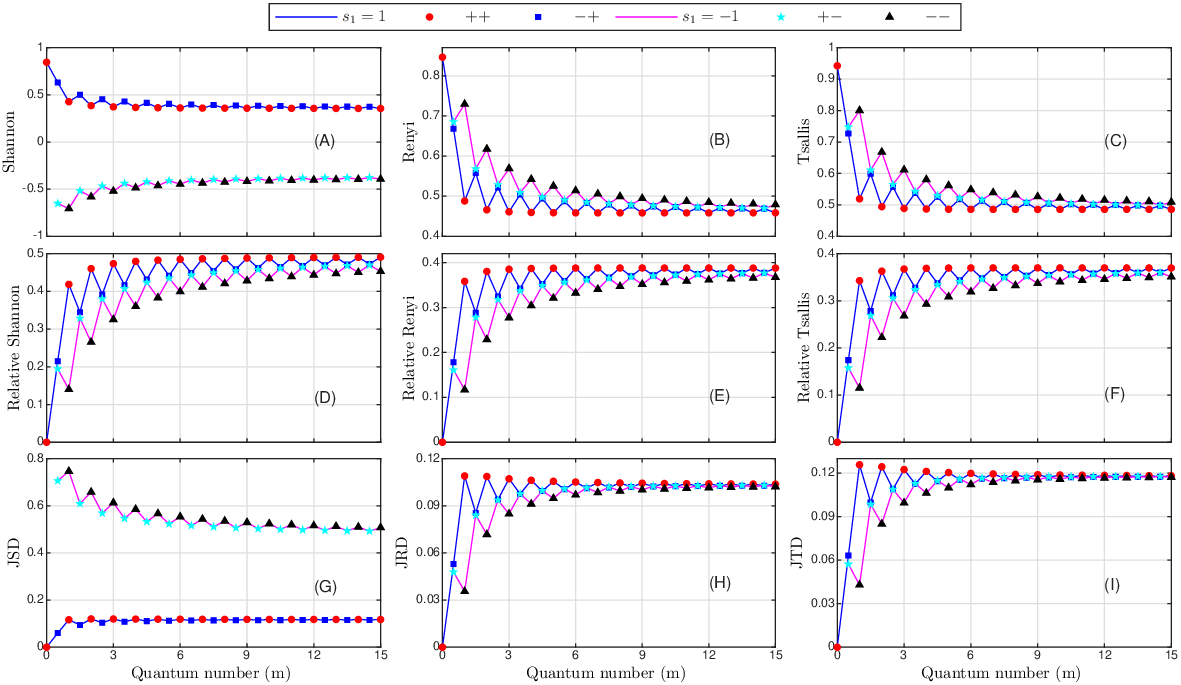}
	\caption{\label{fig4.reflection.g}Direct effect of $\widehat{R}_x, \widehat{R}_y$ on various measures for the $\phi$-function, $G_m^{(s_1,s_2)}(\phi)$ with respect to $m$ for $\hbar=\mu=1,\omega=1$, $\mu_x=0.45$, $\mu_y=0.4$, $\mu_z=0.3$. The order $\widetilde{\a}=0.75$ is used for (B), (C), (E), (F), (H) and (I).}
\end{figure}

\begin{figure}[t]
	\centering
	\includegraphics[width=18cm,height=12cm]{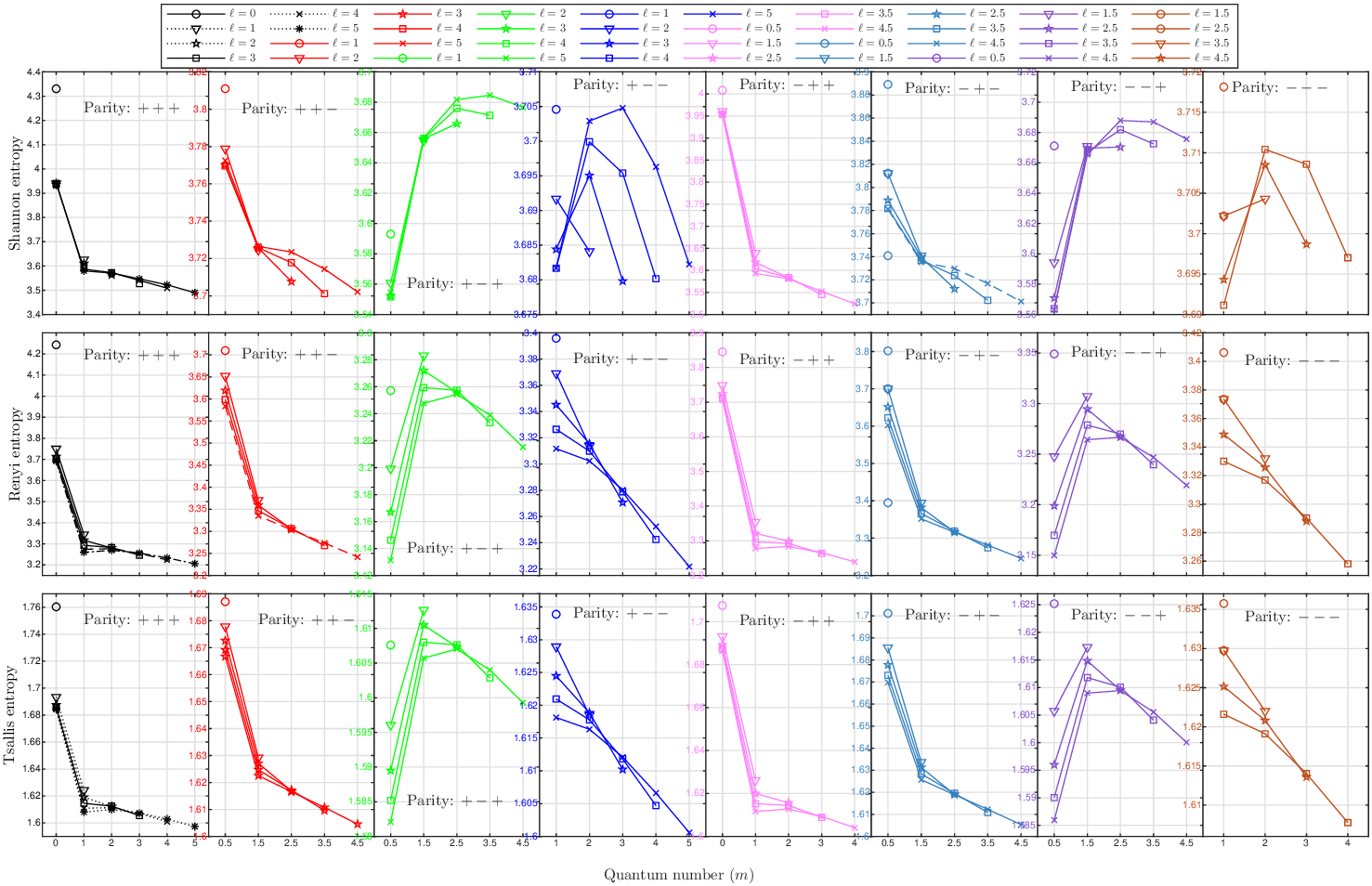}
	\caption{\label{fig5.reflection.total} The effect of reflection operators $\widehat{R}_x, \widehat{R}_y$ on $\mathcal{S}, \mathcal{R}, \mathcal{T}$, w.r.t. $m$ for $\mu_x=0.1, \mu_y=-0.2, \mu_z=0.1$.}
\end{figure}

	The relative R\'enyi information, $R_{rel}^{(\widetilde{\a})}(\rho_m,\rho_n)$ of two density functions, $\rho_{m}$ and $\rho_{n}$ is defined by \cite{jrd.crrao},  
	\beq
	\ds R_{rel}^{(\widetilde{\a})}(\rho_{n,\ell,m}^{(s_1,s_2,s_3)}(\mathbf{r}),\rho_{n',\ell',m'}^{(s_1',s_2',s_3')}(\mathbf{r}))=\ds\f{1}{\widetilde{\a}-1}\ln\left(\int\f{[\rho_{n,\ell,m}^{(s_1,s_2,s_3)}(\mathbf{r})]^{\widetilde{\a}}}{[\rho_{n',\ell',m'}^{(s_1',s_2',s_3')}(\mathbf{r})]^{\widetilde{\a}-1}}d\chi\right).
	\eeq
	The above can be written as a sum of the contributions from marginal densities, as follows: 
	\beq 
	\ds R_{rel}^{(\widetilde{\a})}(\rho_{n,\ell,m}^{(s_1,s_2,s_3)}(\mathbf{r}),\rho_{n',\ell',m'}^{(s_1',s_2',s_3')}(\mathbf{r}))=\ds R_{rel}^{(\widetilde{\a})}\left(\rho_{n,\ell,m}^{(r)},\rho_{n',\ell',m'}^{(r)}\right)+R_{rel}^{(\widetilde{\a})}\left(\rho_{\ell,m}^{(\theta,s_3)},\rho_{\ell',m'}^{(\theta,s_3')}\right)+R_{rel}^{(\widetilde{\a})}\left(\rho_{m}^{(\phi,s_1,s_2)},\rho_{m'}^{(\phi,s_1',s_2')}\right),
	\eeq
	where
	$\ds R_{rel}^{(\widetilde{\a})}\left(\rho_{n,\ell,m}^{(r)},\rho_{n',\ell',m'}^{(r)}\right)=\ds\f{1}{\widetilde{\a}-1}\ln\left(\mathcal{J}_{(n,\ell,m; \, n',\ell',m')}^{(r)}(\widetilde{\a})\right)$, $\ds R_{rel}^{(\widetilde{\a})}\left(\rho_{\ell,m}^{(\theta,s_3)},\rho_{\ell',m'}^{(\theta,s_3')}\right)=\ds\f{1}{\widetilde{\a}-1}\ln\left(\mathcal{J}_{(\ell,m;\, \ell',m')}^{(\theta)}(\widetilde{\a})\right)$ and $\ds R_{rel}^{(\widetilde{\a})}\left(\rho_{m}^{(\phi,s_1,s_2)},\rho_{m'}^{(\phi,s_1',s_2')}\right)=\ds\f{1}{\widetilde{\a}-1}\ln\left(\mathcal{J}_{m,m'}^{(\phi)}(\widetilde{\a})\right)$. 
	Note that $R_{rel}^{(\widetilde{\a})}(\rho,\rho')$ reduces to $S_{KL}\left(\rho,\rho^{'}\right)$ when $\widetilde{\a}\rightarrow1$. Also, $R_{rel}^{(\widetilde{\a})}(\rho,\rho')\ge0$ and $R_{rel}^{(\widetilde{\a})}(\rho,\rho')=0$ if and only if $\rho=\rho'$.
	With reference to ground state with parities $+++$, these entropies, in an arbitrary state, can be written as:  
	$\ds R_{rel}^{(\widetilde{\a})}\left(\rho_{n,\ell,m}^{(r)},\rho_{0,0,0}^{(r)}\right)=\ds\f{1}{\widetilde{\a}-1}\left[\ln\left(\mathcal{J}_{(n,\ell,m; \, 0,0,0)}^{(r)}(\widetilde{\a})\right)\right]$, $\ds R_{rel}^{(\widetilde{\a})}\left(\rho_{\ell,m}^{(\theta,s_3)},\rho_{0,0}^{(\theta,s_3')}\right)=\ds\f{1}{\widetilde{\a}-1}\left[\ln\left(\mathcal{J}_{(\ell,m; \, 0,0)}^{(\theta)}(\widetilde{\a})\right)\right]$ and $\ds R_{rel}^{(\widetilde{\a})}\left(\rho_{m}^{(\phi,s_1,s_2)},\rho_{0}^{(\phi,s_1',s_2')}\right)=\ds\f{1}{\widetilde{\a}-1}\left[\ln\left(\mathcal{J}_{m,0}^{(\phi)}(\widetilde{\a})\right)\right]$. These are obtained for (+)ve integral order $\widetilde{\a}\in\mathbb{N}$.
	The measures corresponding to densities $\rho_{n,\ell,m}^{(r)}$,  $\rho_{\ell,m}^{(\theta,s_3)}$ and $\rho_{m}^{(\phi,s_1,s_2)}$, 
	are displayed in the middle (E) panels of second columns of Figs.~\ref{fig2.reflection.r}, \ref{fig3.reflection.h}, \ref{fig4.reflection.g} w.r.t. $n,\ell, m$ quantum numbers. The fixed parameters used in the calculation, are indicated in the respective figures. 
	
	Next, the generalized Jensen-R\'enyi divergence (GJRD) of order $\widetilde{\a}$ is defined by \cite{jtd}, 
	\beq
	GJRD^{(\widetilde{\a})}(\rho_1,\rho_2,\cdots,\rho_n;\lambda_1,\lambda_2,\cdots,\lambda_n)=\mathcal{R}^{(\widetilde{\a})}\left(\sum_{i=1}^{n}\lambda_i\rho_i\right)-\sum_{i=1}^{n}\lambda_i\mathcal{R}^{(\widetilde{\a})}\left(\rho_i\right)
	\eeq
	whereas $GJRD^{(\widetilde{\a})}(\rho_1,\rho_2;1/2,1/2)=JRD$ can be expressed \cite{jrd.crrao} as, 
	\beq
	\ds JRD^{(\widetilde{\a})}(\rho_{n,\ell,m}^{(s_1,s_2,s_3)}(\mathbf{r}),\rho_{n',\ell',m'}^{(s_1',s_2',s_3')}(\mathbf{r}))=\ds\mathcal{R}^{(\widetilde{\a})}\left(\frac{\rho_{n,\ell,m}^{(s_1,s_2,s_3)}(\mathbf{r})+\rho_{n',\ell',m'}^{(s_1',s_2',s_3')}(\mathbf{r})}{2}\right)-\frac{1}{2}\left[\mathcal{R}^{(\widetilde{\a})}(\rho_{n,\ell,m}^{(s_1,s_2,s_3)}(\mathbf{r}))+\mathcal{R}^{(\widetilde{\a})}(\rho_{n',\ell',m'}^{(s_1',s_2',s_3')}(\mathbf{r}))\right].
	\eeq
	One sees that $JRD(\rho,\rho')\ge0$, $JRD(\rho,\rho')\ge0=JRD(\rho',\rho)\ge0$ and vanishes when $\rho=\rho'$. Moreover, the square root $\sqrt{JRD(\rho,\rho')}$ satisfies the metric properties \cite{gjsd}. The analytical expression of JRD for $\widetilde{\a}\in\mathbb{N}-\{1\}$ is given below, 
	\beq
	\ba{lr}
	JRD^{(\widetilde{\a})}\left(\rho_{n,0,0}^{(s_1,s_2,s_3)}(\mathbf{r}),\rho_{0,0,0}^{(s_1',s_2',s_3')}(\mathbf{r})\right)&=\ds\f{1}{1-\widetilde{\a}}\left[\ln\left\{\sum\limits_{i=0}^{\widetilde{\a}}\binom{\widetilde{\a}}{i}\mathcal{M}_{n,n'} ^{(r,i,\widetilde{\a})}\mathcal{M}_{\ell,\ell'}^{(\theta,i,\widetilde{\a})}\mathcal{M}_{m,m'} ^{(\phi,i,\widetilde{\a})}\right\}-\widetilde{\a}\ln[2]\right]\\
	&\ds-\frac{1}{2}\left[ \mathcal{R}_{n,\ell,m}^{(s_1,s_2,s_3,\widetilde{\a})}+\mathcal{R}_{n',\ell',m'}^{(s_1',s_2',s_3',\widetilde{\a})}\right],
	\ea
	\eeq
	where the following quantities have been defined,  
	\beq 
	\ba{ll}
	\ds\mathcal{M}_{n,n'} ^{(r,i,\widetilde{\a})}= \ds\int_0^{\infty} \left[ \rho_{n,l,m}^{(r)} \right]^i \left[ \rho_{n',l',m'}^{(r)} \right]^{\widetilde{\a} - i}d\chi_r
	=\ds \tilde{\tau}_{n,n'}^{(i)} F_A^{(2\widetilde{\a}+1)} \left( 	\ba{r}
	\tilde{\mu} +1; ~\overbrace{-n,\ \ldots,-n}^{2i},\overbrace{-n',\ldots,-n'}^{2\widetilde{\a}-2i},0;  \\
	\underbrace{L + \frac{3}{2},\ldots,L + \frac{3}{2}}_{2i} \underbrace{L' + \frac{3}{2}, \ldots,L' + \frac{3}{2}}_{2\widetilde{\a}-2i} 1;
	\ea\overbrace{\left\{ \frac{1}{\widetilde{\a}}, \ldots, \frac{1}{\widetilde{\a}} \right\}}^{2\widetilde{\a}},1
	\right),\\
	\ds\mathcal{M}_{\ell,\ell'}^{(\theta,i,\widetilde{\a})}= \ds\int_0^{\pi} \left[ \rho_{\ell,m}^{(\theta,s_3)} \right]^i \left[ \rho_{\ell',m'}^{(\theta,s_3)} \right]^{\widetilde{\a} - i}  \, d\chi_{\theta} =\tilde{\tau}^{(\theta,i,\widetilde{\a})}_{\ell,m,\ell',m'} 2^{	\tilde{\a}_1+	\tilde{\b}_2} \mathcal{B}\left(\tilde{\a}_1+1,\tilde{\b}_1+1\right),\\
	\ds\mathcal{M}_{m,m'} ^{(\phi,i,\widetilde{\a})}= \ds\int_0^{2\pi} \left[ \rho_{m}^{(\phi,s_1,s_2)} \right]^i \left[ \rho_{m'}^{(\phi,s_1,s_2)} \right]^{\widetilde{\a} - i}  \, d\chi_{\phi} =\tilde{\tau}_{m,m'}^{(\phi,i,\widetilde{\a})} 2^{	\tilde{\a'}_2+	\tilde{\b'}_2} \mathcal{B}\left(\tilde{\a'}_2+1,\tilde{\b'}_2+1\right),
	\ea 
	\eeq
	with 
	\beq 
	\ba{ll}
	\tilde{\tau}_{n,n'}^{(i)} &= \ds\frac{[N_n]^{2i} [N_{n'}]^{2\widetilde{\a} - 2i}\Gamma(\mu+1)}{\lam^{\widetilde{\a} + L i + (L' - a)(\widetilde{\a} - 1) + \frac{1}{2}}\widetilde{\a}^{\mu+1}}\binom{n + L + \frac{1}{2}}{n}^{2i}	\binom{n' + L' +\frac{1}{2}}{n'}^{2\widetilde{\a} - 2i}		\left[ \frac{n!}{\left( L + \tfrac{3}{2} \right)_n} \right]^{2i}		\left[ \frac{n'!}{\left( L' + \tfrac{3}{2} \right)_{n'}} \right]^{2\widetilde{\a} - 2i},\\
	\tilde{\mu} &= \widetilde{\a} + L i + L' (\widetilde{\a} - i) - \widetilde{\a}(\widetilde{\a} - 1),
	\ea 
	\eeq 
	\beq
	\ba{lr}
		\tilde{\tau}^{(\theta,i,\widetilde{\a})}_{\ell,m,\ell',m'}&= \ds2 \left[ N^{\theta}_{\ell,m} \right]^{2i} \left[ N^{\theta}_{\ell',m'} \right]^{2\widetilde{\a}- 2i} \left[ \frac{(\g + 1)_\ell}{\ell} \right]^{2i} \left[ \frac{(\g' + 1)_{\ell'}}{\ell'} \right]^{2\widetilde{\a} - 2i} \sum_{k_1 = 0}^{2i\ell} \sum_{k_2 = 0}^{(2\widetilde{\a}-2i)\ell'} \frac{(2i)!}{(2i + k_1)!}  \frac{(2\widetilde{\a} - 2i)!}{(k_2 + 2\widetilde{\a} - 2i)!} \\			
		&\times \ds \widetilde{B}_{2i + k_1, 2i}  \left( C^{(\ell)}_0, 2! C^{(\ell)}_1,..., (k_1+1)! C^{(\ell)}_{k_1} \right)  \widetilde{B}_{2\widetilde{\a} - 2i + k_2, 2\widetilde{\a} - 2i}  \left( C^{(\ell')}_0, 2! C^{(\ell')}_1,..., (k_2+1)! C^{(\ell')}_{k_2}  \right) ,
		\ea
	\eeq
	\beq
		\tilde{\a}_1=2i \sigma + \sigma'(2\widetilde{\a} - 2i) +k_1 + k_2 + \mu_1 + \mu_2,
		\tilde{\b}_1=2\widetilde{\a} \varsigma + \mu_3 - \frac{1}{2} ,
	\eeq
	\beq
	\ba{lr}
	\tilde{\tau}_{m,m'}^{(\phi,i,\widetilde{\a})}&= 4\ds  \left[ N^{\phi}_{m} \right]^{2i} \left[ N^{\phi}_{m'} \right]^{2\widetilde{\a} - 2i} \left[ \frac{(\a + 1)_m}{m} \right]^{2\widetilde{\a}}  \sum\limits_{k_1 = 0}^{2im} \sum_{k_2 = 0}^{(2\widetilde{\a}-2i)m'} \left[\frac{(2i)! \widetilde{B}_{2i + k_1, 2i}  \left( C^{(m)}_0, 2! C^{(m)}_1,..., (k_1+1)! C^{(m)}_{k_1} \right)}{(2i + k_1)!}  \right]\\			
	&\ds\times\left[\frac{(2\widetilde{\a} - 2i)!\, \widetilde{B}_{2\widetilde{\a} - 2i + k_2, 2\widetilde{\a} - 2i}  \left( C^{(m')}_0, 2! C^{(m')}_1,..., (k_2+1)! C^{(m')}_{k_2} \right)}{(k_2 + 2\widetilde{\a} - 2i)!}  \right], 
	\ea
	\eeq
	and
	\beq
	\tilde{\a'}_2= 2\alpha \nu +k_1 + k_2 + \mu_1 + \mu_2,
	\tilde{\b'}_2=2\alpha \xi - \frac{1}{2} .
	\eeq
	Here we have employed Lauricella function, $F_A^{(s)}$ of $s$ variables, with $2s+1$ parameters \cite{srivastava1985,srivastava2003}, given by, 
	\beq
	\ds F_{A}^{(s)}\left(\ba{r}a; \overbrace{b_1, \ldots, b_s}; \\\underbrace{c_1, \ldots, c_s}; \ea\overbrace{x_1, \ldots, x_s}\right)
	= \ds\sum_{j_1, \ldots, j_s = 0}^{\infty} \frac{(a)_{j_1 + \cdots + j_s} (b_1)_{j_1} \cdots (b_s)_{j_s}}{(c_1)_{j_1} \cdots (c_s)_{j_s}} \cdot \frac{x_1^{j_1} \cdots x_s^{j_s}}{j_1! \cdots j_s!}.
	\eeq
	Similarly, analytical values of JRD of marginal density functions are given by, 
		\beq
		\ba{ll} 
	\ds JRD^{(\widetilde{\a})}\left(\rho_{n,\ell,m}^{(r)},\rho_{n',\ell',m'}^{(r)}\right)=\ds\f{1}{1-\widetilde{\a}}\left[\ln\left(\sum\limits_{i=0}^{\widetilde{\a}}\binom{\widetilde{\a}}{i}\mathcal{M}_{n,n'}^{(r,i,\widetilde{\a})}\right)-\widetilde{\a}\ln[2]\right]-\frac{1}{2}\left[\mathcal{R}_{n,\ell,m}^{(r,\widetilde{\a})}+\mathcal{R}_{n',\ell',m'}^{(r,\widetilde{\a})}\right],\\
	\ds JRD^{(\widetilde{\a})}\left(\rho_{\ell,m}^{(\theta,s_3)},\rho_{\ell',m'}^{(\theta,s_3)}\right)=\ds\f{1}{1-\widetilde{\a}}\left[\ln\left(\sum\limits_{i=0}^{\widetilde{\a}}\binom{\widetilde{\a}}{i}\mathcal{M}_{\ell,\ell'}^{(\theta,i,\widetilde{\a})}\right)-\widetilde{\a}\ln[2]\right]-\frac{1}{2}\left[\mathcal{R}_{\ell,m}^{(\theta,s_3,\widetilde{\a})}+\mathcal{R}_{\ell',m'}^{(\theta,s_3',\widetilde{\a})}\right],\\
	\ds JRD^{(\widetilde{\a})}\left(\rho_{m'}^{(\phi,s_1,s_2)},\rho_{m'}^{(\phi,s_1,s_2)}\right)=\ds\f{1}{1-\widetilde{\a}}\left[\ln\left(\sum\limits_{i=0}^{\widetilde{\a}}\binom{\widetilde{\a}}{i}\mathcal{M}_{m,m'}^{(\phi,i,\widetilde{\a})}\right)-\widetilde{\a}\ln[2]\right]-\frac{1}{2}\left[\mathcal{R}_m^{(\phi,s_1,s_2,\widetilde{\a})}+\mathcal{R}_{m'}^{(\phi,s_1',s_2',\widetilde{\a})}\right],~\widetilde{\a}\in\mathbb{N}.
	\ea 
	\eeq 
	One can find JRD for (+)ve integral order. The numerical JRD corresponding to functions, $\rho_{n,\ell,m}^{(r)}$, $\rho_{\ell,m}^{(\theta,s_3)}$, $\rho_{m}^{(\phi,s_1,s_2)}$ are presented for non-integer positive real order. These are shown in lower middle (H) panels of Figs.~\ref{fig2.reflection.r}, \ref{fig3.reflection.h}, \ref{fig4.reflection.g} w.r.t. 
	$n, \ell, m$ state quantum numbers. The parameters used in the calculation are mentioned in respective figures. 

	\subsubsection{Relative Tsallis and generalized Jensen-Tsallis Divergence (GJTD)}
	The general form of relative Tsallis entropy corresponding to two density functions $\rho_{m}$ and $\rho_n$ \cite{relative.tsallis,relative.tsallis2,relative.tsallis3} is, 
	\beq
	\ba{ll}
	\ds T_{\rho,\rho'}^{(\widetilde{\a})}=T_{rel}^{(\widetilde{\a})}\binom{n,\ell,m;s_1,s_2,s_3}{n',\ell',m';s_1',s_2',s_3'}&
	=\ds-\int\rho_{n,\ell,m}^{(s_1,s_2,s_3)}(\mathbf{r})\ln_{\widetilde{\a}}\left[\frac{\rho_{n',\ell',m'}^{(s_1',s_2',s_3')}(\mathbf{r})}{\rho_{n,\ell,m}^{(s_1,s_2,s_3)}(\mathbf{r})}\right]d\chi.
	\ea
	\eeq
	It satisfies the pseudo-additive property, allowing one to write, 
	\beq
	\ba{ll}
	T_{\rho,\rho'}^{(\widetilde{\a})}&=\ds T_{R,R'}^{(r,\widetilde{\a})}+T_{H,H'}^{(\theta,\widetilde{\a})}+T_{G,G'}^{(\phi,\widetilde{\a})}
	+(1-\widetilde{\a})\left[T_{R,R'}^{(r,\widetilde{\a})}T_{H,H'}^{(\theta,\widetilde{\a})}+T_{H,H'}^{(\theta,\widetilde{\a})}T_{G,G'}^{(\phi,\widetilde{\a})}+T_{G,G'}^{(\phi,\widetilde{\a})}T_{R,R'}^{(r,\widetilde{\a})}\right] +(1-\widetilde{\a})^2T_{R,R'}^{(r,\widetilde{\a})}T_{H,H'}^{(\theta,\widetilde{\a})}T_{G,G'}^{(\phi,\widetilde{\a})}, 
	\ea 
	\eeq
	where \\
	$T_{R,R'}^{(r,\widetilde{\a})}=T_{R,R'}^{(r,\widetilde{\a})}\binom{n,\ell,m}{n',\ell',m'}=\ds-\int\rho_{n,\ell,m}^{(r)}\ln_{\widetilde{\a}}\left[\frac{\rho_{n',\ell',m'}^{(r)}}{\rho_{n,\ell,m}^{(r)}}\right]d\chi_r$, $T_{H,H'}^{(\theta,\widetilde{\a})}=T_{H,H'}^{(\theta,\widetilde{\a})}\binom{\ell,m;s_3}{\ell',m';s_3'}=\ds-\int\rho_{\ell,m}^{(\theta,s_3)}\ln_{\widetilde{\a}}\left[\frac{\rho_{\ell',m'}^{(\theta,s_3')}}{\rho_{\ell,m}^{(\theta,s_3)}}\right]d\chi_{\theta}$ and \\
	$T_{G,G'}^{(\phi,\widetilde{\a})}=T_{G,G'}^{(\phi,\widetilde{\a})}\binom{m;s_1,s_2}{m';s_1',s_2'}=\ds-\int\rho_{m}^{(\phi,s_1,s_2)}\ln_{\widetilde{\a}}\left[\frac{\rho_{m'}^{(\phi,s_1',s_2')}}{\rho_{m}^{(\phi,s_1,s_2)}}\right]d\chi_{\phi}$, are the corresponding entropies of marginal density functions. Additionally, it has properties such as positivity, concavity and bounded-ness \cite{borland1998,abe2003_pla,abe2003_pra,rel.tsallis}. 
	
	One can observe that: 
	\begin{equation*}
			\begin{array}{ll}
				(i)~T_{\rho,\rho'}^{(\widetilde{\a})},T_{R,R'}^{(r,\widetilde{\a})}, T_{H,H'}^{(\theta,\widetilde{\a})}, T_{G,G'}^{(\phi,\widetilde{\a})}\ge 0,\mbox{~for~} 0<\widetilde{\a}<1,\\
				(ii)~T_{\rho,\rho'}^{(\widetilde{\a})} \le S_{KL}\left(\rho,\rho'\right) \le T^{(2-\widetilde{\a})}_{\rho,\rho'}, \mbox{~for~} 0\le \widetilde{\a}<1,\\
				(ii)~T^{(2-\widetilde{\a})} \le S_{KL} \le T^{(\widetilde{\a})} ,\mbox{~for~} 0< \widetilde{\a} \le 1,\\
				(iv) \ds\lim_{\widetilde{\alpha} \to 1} T^{(\widetilde{\a})}{\rho,\rho'} = S_{KL} \left(\rho,\rho'\right).
			\end{array}
		\end{equation*} 
		In addition, for discrete distribution $T^{(\widetilde{\a})}$ satisfies the joint convexity, strong additivity, symmetry and possibility of extension. Moreover, it is $\widetilde{\a}$-analog to relative entropy \cite{abe1997,abe1998}, and jointly convex. 
	The analytical forms can be worked out as:    
	$T_{R,R'}^{(r,\widetilde{\a})}\binom{n,\ell,m}{0,0,0}=\ds\frac{1}{\widetilde{\a}-1}\left[\mathcal{J}_{n,0}^{(r)}(\widetilde{\a})-1\right]$, $T_{H,H'}^{(\theta,\widetilde{\a})}\binom{\ell,m;s_3}{0,0;s_3'}=\ds\frac{1}{\widetilde{\a}-1}\left[\mathcal{J}_{\ell,0}^{(\theta)}(\widetilde{\a})-1\right]$ and $T_{G,G'}^{(\phi,\widetilde{\a})}\binom{m;s_1,s_2}{0;s_1',s_2'}=\ds\frac{1}{\widetilde{\a}-1}\left[\mathcal{J}_{m,0}^{(\phi)}(\widetilde{\a})-1\right]$.
	Finally, the relative Tsallis entropy of joint density function can be written as, 
	\beq
	T_{rel}^{(\widetilde{\a})}\binom{n,\ell,m;s_1,s_2,s_3}{0,0,0;s_1',s_2',s_3'}=\ds\frac{1}{\widetilde{\a}-1}\left[\mathcal{J}_{n,0}^{(r)}(\widetilde{\a})\mathcal{J}_{\ell,0}^{(\theta)}(\widetilde{\a})\mathcal{J}_{m,0}^{(\phi)}(\widetilde{\a})-1\right]. 
	\eeq 
	For marginal densities, $\rho_{n,\ell,m}^{(r)}$, $\rho_{\ell,m}^{(\theta,s_3)}, \rho_{m}^{(\phi,s_1,s_2)}$, these entropies 
	are displayed in third right column middle panels (F) of Figs.~\ref{fig2.reflection.r}, \ref{fig3.reflection.h}, \ref{fig4.reflection.g} w.r.t. $n,\ell,m$ quantum numbers, having the parameters as mentioned in corresponding figures.
	
	Now, the generalized Jensen-Tsallis divergence (GJTD) of order $\widetilde{\a}$ is defined by \cite{jtd}, 
		\beq
		GJTD^{(\widetilde{\a})}(\rho_1,\rho_2,\cdots,\rho_n;\lambda_1,\lambda_2,\cdots,\lambda_n)=\mathcal{T}^{(\widetilde{\a})}\left(\sum_{i=1}^{n}\lambda_i\rho_i\right)-\sum_{i=1}^{n}\lambda_i\mathcal{T}^{(\widetilde{\a})}\left(\rho_i\right)
		\eeq
	The JTD of order $\widetilde{\a}$ can be written as \cite{jtd,jtd2}, 
	\beq
	\ba{ll}
	\ds JTD^{(\widetilde{\a})}(\rho_{n,\ell,m}^{(s_1,s_2,s_3)}(\mathbf{r}),\rho_{n',\ell',m'}^{(s_1',s_2',s_3')}(\mathbf{r}))=\ds\mathcal{T}^{(\widetilde{\a})}\left(\frac{\rho_{n,\ell,m}^{(s_1,s_2,s_3)}(\mathbf{r})+\rho_{n',\ell',m'}^{(s_1',s_2',s_3')}(\mathbf{r})}{2}\right)-\frac{1}{2}\left[\mathcal{T}^{(\widetilde{\a})}(\rho_{n,\ell,m}^{(s_1,s_2,s_3)}(\mathbf{r}))+\mathcal{T}^{(\widetilde{\a})}(\rho_{n',\ell',m'}^{(s_1',s_2',s_3')}(\mathbf{r}))\right].
	\ea
	\eeq
	It may be noted that, both $JRD^{(\widetilde{\a})}(\rho,\rho')$ and $JTD^{(\widetilde{\a})}(\rho,\rho')$ are reduced to $JSD(\rho,\rho')$ when $\widetilde{\a}\rightarrow1$. Furthermore, $JTD^{(\widetilde{\a})}(\rho,\rho')$ is symmetric and non-negative.
	It can be simplified as, 
		\beq 
		\ds JTD^{(\widetilde{\a})}\left(\rho_{n,\ell,m}^{(s_1,s_2,s_3)}(\mathbf{r}),\rho_{n',\ell',m'}^{(s_1',s_2',s_3')}(\mathbf{r})\right)=\ds\frac{1}{\widetilde{\a}-1}\left[\frac{1}{2^{\widetilde{\a}}}\sum\limits_{i=0}^{\widetilde{\a}}\binom{\widetilde{\a}}{i}\mathcal{M}_{n,n'} ^{(r,i,\widetilde{\a})}\mathcal{M}_{\ell,\ell'}^{(\theta,i,\widetilde{\a})}\mathcal{M}_{m,m'} ^{(\phi,i,\widetilde{\a})}-1\right]-\frac{1}{2}\left[ \mathcal{T}_{n,\ell,m}^{(s_1,s_2,s_3,\widetilde{\a})}+\mathcal{T}_{n',\ell',m'}^{(s_1',s_2',s_3',\widetilde{\a})}\right],
		\eeq 
	where $\widetilde{\a}\in\mathbb{N}. $ Similarly we can define JTD of marginal density functions; calculated values are displayed in bottom right panels (I) of 
	Figs.~\ref{fig2.reflection.r}, \ref{fig3.reflection.h}, \ref{fig4.reflection.g} w.r.t. $n,\ell,m$ quantum numbers. The parameters employed are given in figures. 
	
\section{Conclusions}\label{sec5.con}
\emph{Exact} solution is obtained for radial and angular components of DSE for an isotropic harmonic oscillator. The dependence of $r$-, $\theta$- $\phi$-solutions on reflection operators is discussed in detail. The obtained solutions are deployed to pursue various information theoretic measures, such as, Shannon entropy, R\'enyi entropy and Tsallis entropy in exact/quasi-exact manner. In absence of the Dunkl operator, Shannon entropy in position space has been reported earlier \cite{jsd1994}. Here we invoke the concept of factorization method to get the entropy in a DS system \cite{pra1985,jsd1994,jsd2010,jsd2010.jacobi,jsd2011}, which is otherwise, an open problem \cite{arxiv.dn,dn.jmc2023}. Next we proceed for relative information theoretic measures, like relative entropy, relative R\'enyi and relative Tsallis of two density functions, in analytical forms. Lastly, we consider JSD, JRD and JTD analytically for two density functions. To realize the effect of Dunkl parameter we have produced the entropic moments and all the above measures for different parities, in graphical manner. 

\subsubsection*{Acknowledgement}
AKR gratefully acknowledges financial support from DST SERB (sanction order CRG/2023/004463), and AH from CSIR, New Delhi (09/0921(16264)/2023-EMR-I), for SRF. 
\small{
	
}
\end{document}